\documentclass[aps,pre,onecolumn,superscriptaddress]{revtex4}
\usepackage{amsfonts}
\usepackage{epsfig}
\usepackage{amsmath,amssymb}
\usepackage{times}
\usepackage{xcolor}
\usepackage{multirow}

\begin{document}

\title{Impacts of complex behavioral responses on asymmetric interacting spreading dynamics in multiplex networks}

\date{\today}

\author{Quan-Hui Liu}
\affiliation{Web Sciences Center, University of Electronic Science and
Technology of China, Chengdu 611731, China}
\affiliation{Big Data Research Center, University of Electronic Science and
Technology of China, Chengdu 611731, China}

\author{Wei Wang}
\affiliation{Web Sciences Center, University of Electronic Science and
Technology of China, Chengdu 611731, China}
\affiliation{Big Data Research Center, University of Electronic Science and
Technology of China, Chengdu 611731, China}

\author{Ming Tang\footnote{Correspondence to: tangminghan007@gmail.com}}
\affiliation{Web Sciences Center, University of Electronic Science and
Technology of China, Chengdu 611731, China}
\affiliation{Big Data Research Center, University of Electronic Science and
Technology of China, Chengdu 611731, China}

\author{Hai-Feng Zhang\footnote{Correspondence to: haifengzhang1978@gmail.com}}
\affiliation{School of Mathematical Science, Anhui University,
Hefei 230039, China}

\begin{abstract}

Information diffusion and disease spreading in communication-contact layered network are typically asymmetrically coupled with each other, in which disease spreading can be significantly affected by the way an individual being aware of disease responds to the disease. Many recent studies have demonstrated that human behavioral adoption is a complex and non-Markovian process, where the probability of behavior adoption is dependent on the cumulative times of information received and the social reinforcement effect of the cumulative information. In this paper, the impacts of such a non-Markovian vaccination adoption behavior on the epidemic dynamics and the control effects are explored. It is found that this complex adoption behavior in the communication layer can significantly enhance the epidemic threshold and reduce the final infection rate. By defining the social cost as the total cost of vaccination and treatment, it can be seen that there exists an optimal social reinforcement effect and optimal information transmission rate allowing the minimal social cost. Moreover, a mean-field theory is developed to verify the correctness of simulation results.

\end{abstract}

\maketitle

When a disease suddenly emerges, the dynamical processes of disease~\cite{Anderson:1992,Hethcote:2000,Daley:2001,Pastor-Satorras:2001,Newman:2002,Moreno:2002,Romualdo:2015} and information~\cite{Zanette:2002,Liu:2003,Noh:2004} spreading are typically asymmetrically coupled with each other~\cite{Kiss:2010,Sahneh:2012,Wu:2012,Ruan:2012,Jo:2006}. In particular, the spread of a disease can enhance the crisis awareness and thus facilitates the diffusion of the information about the disease~\cite{Funk:2010JTB}. Meanwhile, the diffusion of the information promotes more people to take preventive measures and consequently suppresses the epidemic spreading~\cite{Ruan:2012}. To understand the asymmetric interplay between the two kinds of spreading dynamics is of great importance for predicting and controlling epidemics, leading to a new direction of research in complex network science~\cite{Funk:2009,Granell:2013,Wei:2014}. Funk \emph{et al.} first presented an epidemiological model by incorporating the spread of awareness in a well-mixed population, and found that the awareness-based response can markedly reduce the final infection rate. When the awareness is sufficiently strong so as to modify the key parameters associated with the spreading dynamics such as the infection and recovery rates, the epidemic threshold can be enhanced~\cite{Funk:2009}. Ruan \emph{et al.} studied a susceptible-infected-recovered (SIR) model with information-driven vaccination, and found the epidemic spreading can be significantly suppressed when the information is well spread~\cite{Ruan:2012}.

With the development of technology, the information about disease can quickly diffuse through different channels, such as the word of mouth, news media and online social networks. Usually, the pathways for information spreading are different from the pathways for disease spreading. In view of this, the asymmetric interplay between the information and the epidemic spreading dynamics needs to be considered within multiplex network framework~\cite{Boccaletti:2014,Wei:2014,Salehi:2014,Kivel:2014,Granell:2013,Kim:2013}. In a multiplex network (multilayer network or overlay network), each network layer for one type of transportation process has an identical set of nodes and a distinct internal structure. And the interplay between multiple layers has diverse characteristics, such as inter-similarity~\cite{Parshani:2010}, multiple support dependence~\cite{Shao:2011}, and inter degree-degree correlation~\cite{Lee:2012}, etc. Along this line, Granell \emph{et~al.} established a two susceptible-infected-susceptible (SIS) processes coupled model to investigate the inhibitory effect of awareness spreading on epidemic spreading dynamics in a multiplex network, and the results showed that the epidemic threshold was determined by the structures of the two respective networks as well as the effective transmission rate of awareness~\cite{Granell:2013}. Wang \emph{et~al.} studied the asymmetrically interacting spreading dynamics based on a two susceptible-infected-recovered (SIR) processes coupled model in multiplex networks, and found that the outbreak of disease can lead to the propagation of information, and rise of epidemic threshold~\cite{Wei:2014}.

In the asymmetrically interacting spreading dynamics, how an individual being aware of disease responds to the disease can significantly affect the epidemic spreading~\cite{Ruan:2012,Zhang:2014,Wu:2012}. Sahneh~\emph{et al.} introduced an \emph{alter} state into the SIS model, where the alerted individuals sensing infection adopt a preventive behavior. When the preventive behavior is implemented timely and effectively, disease cannot survive in the long run and will be completely contained~\cite{Sahneh:2012}. Zhang \emph{et al.} investigated to what extent behavioral responses based on local infection information can affect typical epidemic dynamics, and found that such responses can augment significantly the epidemic threshold, regardless of SIS or SIR processes~\cite{Zhang:2014}. All of the previous studies were built on a basic assumption: the behavioral responses to the disease, which is a Markovian process without memory, depend only on \emph{current} dynamical information such as infected neighbors.

However, behavioral response or behavior adoption is not a simple Markovian process which depends only on current dynamical information. Recent researches on behavior adoption such as innovation~\cite{Young:2011} and healthy activities~\cite{Centola:2011} have confirmed that the adoption probability is also affected by \emph{previous} dynamical information. This is equivalent to social affirmation or reinforcement effect, since multiple confirmation of the credibility and legitimacy of the behavior are always sought~\cite{Centola:2010,Dodds:2004,Dodds:2005,Weiss:2014,Centola:2007}. Specifically for an individual, if some of his/her friends have adopted a particular behavior before a given time whereas the other friends newly adopt the behavior, whether he/she adopt the behavior will take all the adopted friends' adoption into account. Taking the adoption of healthy behavior as an example, Centola has demonstrated that the probability for an individual to adopt a healthy behavior depends on the times of being informed~\cite{Centola:2010}; in the microblogging retweeting process, the authors have shown that the probability of one individual retweeting a message increases when more friends have retweeted the message~\cite{Zhang:2013,Hodas:2014}. Based on the memory of previous information, this reinforcement effect makes the behavior adoption processes essentially non-Markovian and more complicated.

As we know, taking vaccination against disease may carry some side effects or certain cost~\cite{Altarelli:2014,ZhangHF:2014}, so the decision to take vaccination is worth pondering. Before taking a certain vaccine, people need to confirm the correctness of information which usually relies on the cumulative times of received information and the social reinforcement effect. Thus, the adoption of vaccination can be viewed as a complex adoption behavior. In this paper, the impact of complex vaccination adoption behavior on the two interacting spreading dynamics in a double-layer network is investigated. It is assumed that in physical-contact layer, the probability for an individual to adopt vaccination is determined by the times of the information about disease received in the communication layer and the social reinforcement effect of the cumulative information. It is showed by our findings that the two interacting spreading dynamics is remarkably influenced by this complex adoption behavior. In addition, given that taking vaccination as well as treating infected individuals bear certain costs, we define the social cost as the total cost of vaccination and treatment for infected individuals. Then, the effect of this complex vaccination adoption behavior on social cost is explored, and it is found that there are an optimal social reinforcement effect and optimal information transmission rate which entail the minimal social cost.

\section*{Results}
To present our primary study results, we first described the model of multiplex network, the spreading dynamical process in each layer, and the asymmetric interplay between the two spreading processes. Then, we elaborated the theoretical analysis of the asymmetric interacting spreading dynamics in multiplex networks. Finally, we demonstrated the simulation results which are verified by the proposed theory.

\textbf{Model of multiplex network}. A multiplex network with two layers is constructed to represent the contact-communication coupled network. At the beginning, a communication network (labelled $A$) and a contact network (labelled $B$) are respectively generated. Supposing that the degree distribution and network size of communication network $A$ are of $P_A(k_A)$ and $N$ respectively, a random configuration network can be generated according to the given degree distribution, where self-loops or repeated links between a pair of nodes are not
allowed~\cite{Newman:2002}. Meanwhile, layer $B$ is generated in the same way that the network size and degree distribution are given as $N$ and $P_B(k_B)$, respectively. After that, each node of layer $A$ is matched one-to-one with that of layer $B$ randomly. Moreover, to facilitate the analysis, the constructed double-layer network is an uncorrelated double-layer network, and the joint probability distribution of degree $k_A$ and degree $k_B$ of the same node can be written as $P_{AB}(k_A,k_B)$ = $P_A(k_A)$$P_B(k_B)$. It means that the degree distribution of one layer is independent of that of the other layer completely. In addition, when the network is very large and sparse, links in the double layers are scarcely overlapped due to random linking in random configuration network model. The theoretical framework of the asymmetric interacting spreading processes in this paper can be easily generalized to the multiplex networks with inter-layer degree correlations~\cite{Wei:2014} and overlapping links~\cite{Marceau:2011}.

\textbf{Two interacting spreading dynamical processes}.
In such a double-layer network, an infectious disease spreads through physical contact layer (layer $B$), and the triggered information about the disease diffuses through a communication layer (layer $A$). In the communication layer (layer $A$), an improved susceptible-infected-recovered (SIR) model~\cite{Moreno:2002} is used to describe the dissemination of information about the disease. In this model, each node can be in one of the following three states: (1) susceptible state ($S$) in which the node has not received any information about the disease; (2) informed state ($I$), where the node has received the information at least one time and is capable of transmitting the information to other nodes in the same layer. More importantly, let \textbf{$M$} be the cumulative pieces of information that the node has received from its neighbors, which is used to characterize the memory effect of vaccination adoption behavior~\cite{Dodds:2004, Wang:2015}; and (3) refractory state ($R$), in which the node has received the information but is not willing to pass it on to other nodes. During the process of transmission, each informed node ($I$ state) passes the information to all its neighbors in the communication network $A$ at each time step. If a neighbor is in the $S$ state, it will enter $I$ state and update $M=1$ with probability $\beta_A$. If a neighbor is in the $I$ state, it will receive the information again and update $M = M + 1$ with probability $\beta_A$. Meanwhile, the informed node enters the $R$ state with probability $\mu_A$, and once the node enters the $R$ state, it will keep in this state forever. Furthermore, a node in layer $A$ will get the information about the disease and update $M =1$, once its counterpart node in layer $B$ is infected. As a result, the dissemination of the information over layer $A$ is facilitated by disease transmission in layer $B$.

\begin{figure}
\begin{center}
\includegraphics[width=6.5in]{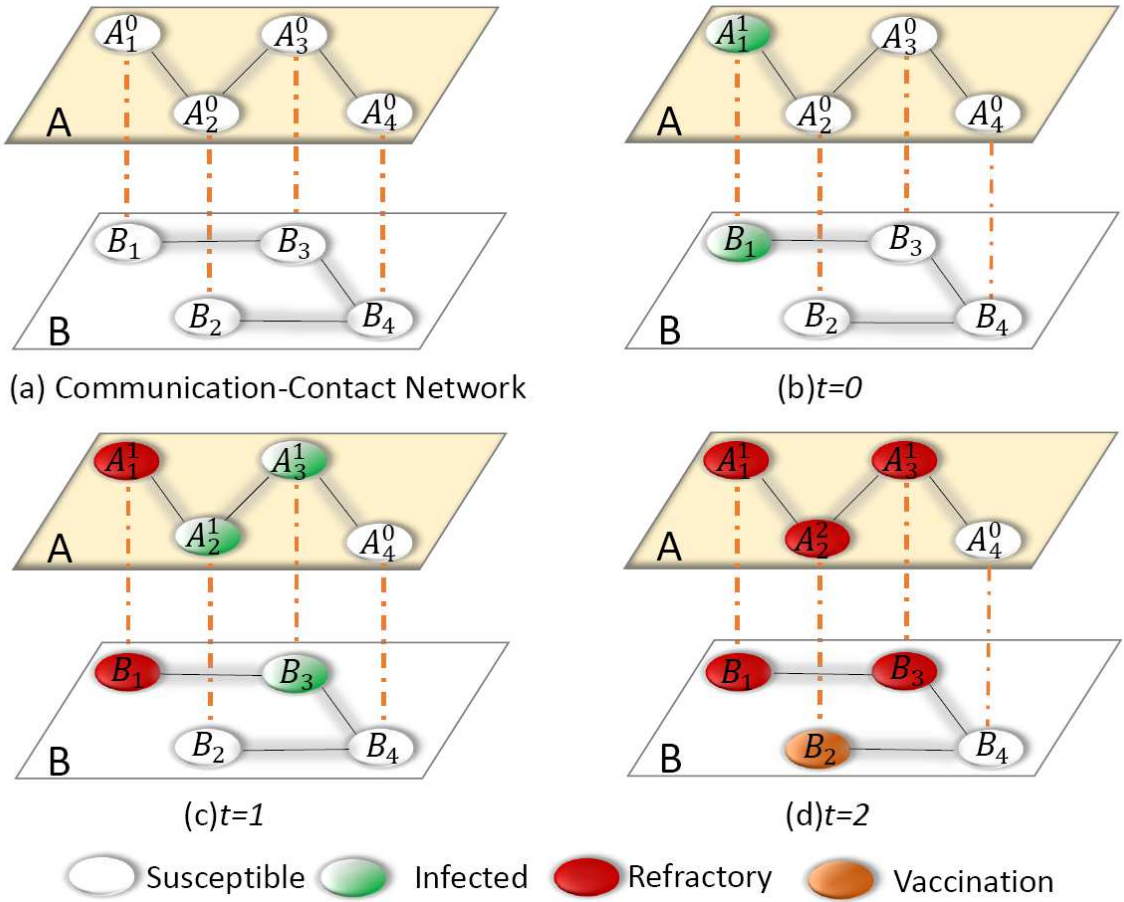}
\caption{\textbf{Illustration of asymmetrically coupled spreading processes in a double-layered communication-contact network.} (a) Communication and contact networks, denoted respectively as layer $A$ and layer $B$, each have four nodes. Each node of layer $A$ is matched one-to-one with that of layer $B$ randomly. A node $i$ in layer $A$ is represented as ${A_i^M}$, where the subscript and superscript respectively represent the index of node and the times of received information. (b) At $t=0$, node $B_1$ in layer $B$ is randomly chosen as the initial infected node and its counterpart, node $A_1$ in layer $A$, gains the information and becomes informed state and updates $M=1$. While all other pairs of nodes, one from layer $A$ and another from layer $B$, are in the susceptible state. (c) At $t=1$, node $B_3$ in layer $B$ can be infected by infected neighbor $B_1$ with probability $\beta_B$, and if it is indeed infected, its corresponding node $A_3$ in layer $A$ will get the information as well and update $M=1$. Within layer $A$ the information is transmitted from $A_1$ to $A_2$, with $M=1$ for $A_2$. Since, by this time, $A_2$ is already aware of the infection spreading, whereas its counterpart $B_2$ in layer $B$ takes vaccination with probability $\xi_1$, but fails. At the same time, node $A_1$ in layer $A$ and its counterpart $B_1$ in layer $B$ enter into the refractory state with probability $\mu_A$ and $\mu_B$, respectively. (d) At $t=2$, in layer $A$, $A_3$ successfully transmits the information to $A_2$. In this case, node $A_2$ updates $M=2$. At the same time, its counterpart $B_2$ in layer $B$ takes vaccination with probability $\xi_2$ and successfully becomes a vaccinated node. The spreading dynamics terminate as all infected/informed nodes have entered into the refractory state.}
\label{fig1}
\end{center}
\end{figure}

The dynamics of epidemic in the contact network $B$ is illustrated by a susceptible-infected-recovery-vaccinated (SIRV) model~\cite{Ruan:2012}, in which a fourth state, the state of vaccination is incorporated into the classical SIR model. The reaction process of the SIR component in layer $B$ is the same as that of the classical SIR model with transmission rate $\beta_B$ and recovery rate $\mu_B$. Since the behavior of taking vaccination against disease is essentially non-Markovian and complicated, we assume that the probability of a susceptible node turning into vaccinated state in layer $B$ depends on the cumulative times of received information (i.e $M$) in layer $A$ and the social reinforcement effect. For a susceptible node in layer $B$, if he receives at least one piece of information at the $t$th time step and has received $M$ times of the information until time $t$, the probability that he takes vaccination at time $t$ will be
\begin{eqnarray}\label{A1}
\xi_{M}=\xi_1+(1-\xi_1)[1-e^{-\alpha(M-1)}],
\end{eqnarray}
where $\xi_1$ is the vaccination adoption probability when a node receives the information about disease for the first time. And $\alpha$ means the node's sensitivity to information, which is used to characterize the strength of social reinforcement effect. When $\alpha>0$, the adoption probability $\xi_M$ increases with the value of $M$. The memory reinforcement effect disappears once $\alpha=0$. For a fixed $M$, the greater value of $\alpha$, the stronger the reinforcement effect (i. e., the greater adoption probability $\xi_M$). As the adoption of vaccination is determined by the cumulative pieces of received information $M$ and the sensitivity factor of social reinforcement effect $\alpha$, it is a typical complex adoption behavior. Our main purpose is to investigate the impact of sensitivity factor $\alpha$ on the two interacting epidemic dynamics. The two spreading processes and their dynamical interplay are schematically illustrated in Fig.~\ref{fig1}. To simplify our descriptions and differentiate the states of nodes in the two layers, $S_A$ ($R_A$) and $S_B$ ($R_B$) are defined to be a node in $S$ ($R$) state in layer $A$ and layer $B$, respectively. Similarly, $I_A$ and $I_B$ are set as nodes in \emph{informed} state and \emph{infected} state in layer $A$ and $B$, respectively. And $V_B$ is the node in \emph{vaccinated} state in layer $B$.

\textbf{Theoretical analysis}. The epidemic threshold and the final infection density are the two key quantities in the dynamics of spreading. Thus, in this paper, a theory is proposed to predict these quantities for both information and epidemic spreading in the double-layer network.

Let $P_A(k_A)$ [$P_B(k_B)$] be the degree distribution of communication layer $A$ (contact layer $B$), and the average degrees of $A$ and $B$ are $\langle k_A\rangle=\sum_{k_A}k_AP_A(k_A)$ and $\langle k_B\rangle=\sum_{k_B}k_BP_B(k_B)$, respectively. Here, our sole focus is the uncorrelated double-layer network, where the joint probability distribution of degree $k_A$ and degree $k_B$ of a node can be expressed as $P_{AB}{(k_A,k_B)}=P_A{(k_A)}P_B{(k_B)}$. Meanwhile, we assume that there is no degree correlations between inner-layer links and inter-layer links. If the specific formula of $P_{AB}{(k_A,k_B)}$ is given, the developed theory can be extended to the correlated double-layer networks~\cite{Wei:2014,Kim:2013, Lee:2012}. The variables of $s_{k_A}^{A}(t)$, $\rho_{k_A}^{A}(t)$ and $r_{k_A}^{A}(t)$ are used to denote the densities of the susceptible, informed, and recovered nodes with degree $k_A$ in layer $A$ at time $t$, respectively. Thereinto, $\rho_{k_A}^{A}(t)=\sum_{m}\rho_{k_A}^{A}(m,t)$, and  $\rho_{k_A}^{A}(m,t)$ is the density of $I_A$ nodes with degree $k_A$ which have received $m$ pieces of information till time $t$. Similarly, $s_{k_B}^{B}(t)$, $\rho_{k_B}^{B}(t)$, $r_{k_B}^{B}(t)$ and $v_{k_B}^{B}(t)$ are the densities of the susceptible, infected, recovered and vaccinated nodes with degree $k_B$ in layer $B$ at time $t$, respectively. The effective transmission rates for the two spreading dynamics are respectively expressed as $\lambda_A=\beta_A/\mu_A$ and $\lambda_B=\beta_B/\mu_B$. Without loss of generality, we set $\mu_A=\mu_B=\mu$, which won't affect the relative sizes of effective information and disease transmission rates.

The mean-field rate equation of the information spreading in layer $A$ is
\begin{eqnarray}
\label{A2}
\frac{ds_{k_A}^A(t)}{dt} & = & -s_{k_A}^A(t){\Psi_{S_A,k_A}^A}(t)-s_{k_A}^A(t)\sum_{k_B}P_B({k_B})\Psi_{S_B,k_B}^B(t),
\end{eqnarray}
where $\Psi_{S_A,k_A}^A(t)$ [$\Psi_{S_B,k_B}^B(t)$] denotes the probability of a $S_A$ ($S_B$) node with degree $k_A$ ($k_B$) in layer $A$ ($B$) being informed (infected) by its neighbor in the same layer at time $t$ (See Methods for details). The first term in the right hand side (RHS) of Eq.~(\ref{A2}) means the loss of $S_A$ nodes since they have received information from their neighbors in layer $A$. And the second term represents the counterpart nodes of $S_A$ nodes in layer $B$ are infected by the disease resulting in the decrease of $S_A$ nodes. For $m=1$, the gain of $\rho_{k_A}^{A}(1,t)$ can only come from $S_{A}$ nodes. But for $m>1$, the density of $\rho_{k_A}^{A}(m,t)$ can be increased by the case in which the $I_A$ nodes have already received $n$ pieces of information and receive $m-n$ pieces of information again at time $t$. As a result, the rate equations of $\rho_{k_A}^{A}(m,t)$ when $m=1$ and $m>1$ should be established, respectively.

When $m=1$, the rate equation of $\rho_{k_A}^{A}(1,t)$ is given as
\begin{eqnarray}
\begin{split}
\label{A3}
\frac{d\rho_{k_A}^A(1,t)}{dt}&=s_{k_A}^A(t)\sum_{n=1}^{k_A}\pi_{S_A,k_A}^A(n)B_{n,1}(\beta_A)\\
&+s_{k_A}^A(t)\sum_{k_B}P_B({k_B})\Psi_{S_B,k_B}^B(t)-\rho_{k_A}^A(1,t)\Psi_{I_A,k_A}^A(t)-\mu\rho_{k_A}^A(1,t),
\end{split}
\end{eqnarray}
where $\pi_{S_A,k_A}^A(n)$ is the probability of a $S_A$ node with degree $k_A$ in layer $A$ which has $n$ ($n\leq{k_A}$) number of informed neighbors, $B_{k,m}(\beta_A)$ denotes the binomial factor $\binom{k}{m}{\beta_A}^m{(1-\beta_A)}^{k-m}$ and $\Psi_{I_A,k_A}^A(t)$ means the probability of an $I_A$ node with degree $k_A$ being informed again by its neighbors in layer $A$ at time $t$ (See Methods for details). The first and second term in the RHS of Eq.~(\ref{A3}) correspond to the case that the $S_A$ node receives one piece of information and the case that the $S_B$ node is infected by the disease, respectively. The third term means that the informed node ($I_A$) which has only received one piece of information previously receives one or more pieces of information at time $t$.
The fourth term describes the recovery of the $I_A$ node.

When $m>1$, the rate equation of $\rho_{k_A}^A(m,t)$ can be described as
\begin{eqnarray}
\label{A4}
\frac{d\rho_{k_A}^A(m,t)}{dt} & = & s_{k_A}^A(t)\sum_{n=m}^{k_A}\pi_{S_A,k_A}^A(n)B_{n,m}(\beta_A)\\
\nonumber & + & \sum_{q=1}^{m-1}\rho_{k_A}^A(q,t)\sum_{n=m-q}^{k_A}\pi_{I_A,k_A}^A(n)B_{n,m-q}(\beta_A)\\
\nonumber & - &\rho_{k_A}^A(m,t)\Psi_{I_A,k_A}^A(t)-\mu\rho_{k_A}^A(m,t),
\end{eqnarray}
where $\pi_{I_A,k_A}^A(n)$ represents the probability of an $I_A$ node with degree $k_A$ to have $n$ ($n\leq{k_A}$) number of informed neighbors (See Methods for details). The first term in the RHS of Eq.~(\ref{A4}) means that a $S_A$ node receives $m$ pieces of information at time $t$. The second term in the RHS of Eq.~(\ref{A4}) denotes the case in which the $I_A$ node with degree $k_A$ has received $q$ ($0<q<m$) pieces of information previously, and then receives $m-q$ pieces of information at time $t$. The third and the fourth term are the same to those of Eq.~(\ref{A3}), which indicate the losses caused by the newly received information and the recovery of $I_A$ to $R_A$, respectively. The rate equation for $r_{k_A}^{A}$ can be written as
\begin{eqnarray}
\label{A5}
\frac{dr_{k_A}^A(t)}{dt} & = & \mu\sum_m\rho_{k_A}^A(m,t).
\end{eqnarray}
The mean-field rate equation of the epidemic spreading in layer $B$ is
\begin{eqnarray}
\begin{split}
\label{A6}
\frac{ds_{k_B}^B(t)}{dt} =-{s_{k_B}^B(t)}\Psi_{S_B,k_B}^B(t)-\sum_{k_A}\chi_{S_A,k_A}^A(t)-s_{k_B}^B(t)\sum_{k_A}\chi_{I_A,k_A}^A(t),
\end{split}
\end{eqnarray}
where $\chi_{S_A,k_A}^A(t)$ [$\chi_{I_A,k_A}^A(t)$] refers to the probability that a $S_A$ ($I_A$) node with degree $k_A$ newly receives information to make its counterpart node in layer $B$ vaccinated (See Methods for details). The first term in the RHS of Eq.~(\ref{A6}) means that the $S_B$ type nodes are infected by their neighbors in layer $B$. The second and third terms in the RHS of Eq.~(\ref{A6}) represent that the $S_B$ nodes' counterpart nodes are respectively in $S_A$ and $I_A$ state in layer $A$, receiving the information about disease and making $S_B$ nodes vaccinated.
\begin{eqnarray}
\label{A7}
\frac{d\rho_{k_B}^B(t)}{dt} & = & s_{k_B}^B(t)\Psi_{S_B,k_B}^B(t)-\mu\rho_{k_B}^B(t),
\end{eqnarray}
\begin{eqnarray}
\label{A8}
\frac{r_{k_B}^B(t)}{dt} & = & \mu\rho_{k_B}^B(t),
\end{eqnarray}
\begin{eqnarray}
\label{A9}
\frac{dv_{k_B}^B(t)}{dt} & = & \sum_{k_A}\chi_{S_A,k_A}^A(t)+s_{k_B}^B(t)\sum_{k_A}\chi_{I_A,k_A}^A(t).
\end{eqnarray}

From Eqs. (2)-(9), the density associated with each distinct state in layer $A$ or $B$ is given by
\begin{eqnarray}
\label{A10}
x^H(t) &=& \sum_{k_H,min}^{k_H,max}P_H({k_H})x_{k_H}^H(t),
\end{eqnarray}
where $H\in\{A,B\}$, $x\in\{s,\rho,r,v\}$, and $k_{H,min}$ ($k_{H,max}$) denotes the smallest (largest) degree of layer $H$.
Specially, the density of $I_A$ node with degree $k_A$ in layer $A$ is $\rho_{k_A}^A(t)=\sum_{m}\rho_{k_A}^A(m,t)$.
The final densities of the whole system can be obtained by taking the limit $t\rightarrow\infty$.

Owing to the complicated interaction between the disease and information spreading process, it is unfeasible to derive the exact threshold values. Thus, a linear approximation method is applied to derive the outbreak threshold of information spreading in layer $A$ (see Supporting Information for details) as
\begin{eqnarray}
\label{A11}
\beta_{Ac} = \left\{\begin{array}{l}\beta_{Au},
\quad for \quad \beta_{B}\leq \beta_{Bu} \\
0, \quad \quad for  \quad \beta_{B}> \beta_{Bu},
\end{array} \right.
\end{eqnarray}
where
\begin{eqnarray}
\beta_{Au}   \equiv   \mu\langle k_A\rangle/({\langle {k_A}^2\rangle
}-\langle k_A\rangle)
\end{eqnarray}
and
\begin{eqnarray}
\beta_{Bu}   \equiv
 \mu\langle k_B\rangle/(\langle {k_B}^2\rangle-\langle k_B\rangle)
\end{eqnarray}
refer to the outbreak threshold of information spreading in layer $A$ when it is isolated from layer $B$, and the outbreak threshold of epidemic spreading in layer $B$ when the coupling
between the two layers is absent, respectively.

For $\beta_A<\beta_{Au}$, Eq.~(\ref{A11}) shows that the information cannot break out in layer $A$ if layer $A$ and layer $B$ are isolated. When the two spreading dynamics are interacting,\emph{ near the epidemic threshold}, the spread of epidemic in layer $B$ can only lead to a few of counterpart nodes in layer $A$ ``infected'' with the information, and thus these informed nodes in layer $A$ have negligible effect on the epidemic dynamics in layer $B$ since $\beta_A<\beta_{Au}$. The above explanation indicates that $\beta_{Bc}\approx\beta_{Bu}$ when $\beta_A<\beta_{Au}$. However, for $\beta_{A}>\beta_{Au}$, the information outbreaks in layer $A$ which makes many counterpart nodes in layer $B$ to be vaccinated, and thus hinders the spread of epidemic in layer $B$. Once a node is in the vaccination state, it will no longer be infected. Usually, we can regard this kind of vaccination as a type of ``disease,'' and every node in layer $B$ can be in one of the two states: infected or vaccinated. Epidemic spreading and vaccination diffusion (derived by information diffusion) can thus be viewed as a pair of competing ``diseases'' spreading in layer $B$~\cite{Karrer:2011}. As pointed out by Karrer and Newman~\cite{Karrer:2011}, when two competing diseases have different growth rates in large size network \emph{N}, they can be treated as if they were in fact spreading non-concurrently, one after the other.

To clarify the interplay between epidemic and vaccination spreading, we should determine which one is the faster ``disease''. At the early stage, the average number of infected and vaccinated nodes in  layer $B$ grows exponentially (see Supporting Information). And the ratio of their growth rate can be expressed as
\begin{eqnarray}
\label{A12}
\theta=\frac{\beta_A\beta_{Bu}}{\beta_B\beta_{Au}}
\end{eqnarray}
When $\theta<1$, \emph{i.e.}, $\beta_B\beta_{Au}>\beta_A\beta_{Bu}$, the disease process grows faster than the vaccination process. In this case, we can ignore the effect of vaccination on epidemic spreading. However, when $\theta>1$, the information process spreads faster than the epidemic process, which is in accord with reality since many on-line social networks and mass media can promote information spreading. Given that vaccination and epidemic can be treated successively and separately, by letting $\beta_B=0$ and obtaining the final density of vaccination $v^B(\infty)|_{\beta_B=0}$ from Eq.~(\ref{A9}), the threshold of epidemic outbreak is given as~\cite{Wei:2014}
\begin{eqnarray}
\label{A13}
{\beta_{Bc}}=\frac{\mu\langle k_B\rangle}{[1-v^B(\infty)|_{\beta_B=0}]({\langle {k_B}^2\rangle}-\langle k_B\rangle)}.
\end{eqnarray}

\textbf{Simulation results}.
The standard configuration model is used to generate a network with power-law degree distribution~\cite{Newman:2005,Catanzaro:2005} for the communication subnetwork (layer A). The contact subnetwork for layer $B$ is of the Erd\H{o}s and R\'{e}nyi (ER) type~\cite{Erdos:1959}. The notation SF-ER is adopted to denote the double-layer network. The network sizes of both layers are set to be $N_A=N_B=10000$ and their average degrees are $\langle k_A\rangle=\langle k_B\rangle=8$. The degree distribution of communication layer $A$ is expressed as $P_A(k_A)=\Gamma{k_A}^{-\gamma_A}$, when the coefficient is $\Gamma=1/\sum_{k_{min}}^{k_{max}}{k_A}^{-\gamma_A}$ and the maximum degree is $k_{max}{\sim}N^{1/(\gamma_A-1)}$. The degree distribution of contact layer $B$ is $P_B(k_B)=e^{-\langle k_B \rangle}{\langle k_B \rangle}^{k_B}/{k_B}!$. Without loss of generality, we set $\gamma_A=3.0$, $\xi_1=0.05$, and \textbf{$\mu_A=\mu_B=\mu=0.5$} in the following simulations. To initiate an epidemic spreading process, a node in layer $B$ is randomly infected and its counterpart node in layer $A$ is thus in the informed state, too. The spreading dynamics terminates when all infected/informed nodes in both layers are recovered, and the final densities $r^A(\infty)$, $r^B(\infty)$, and $v^B(\infty)$ are then recorded. We use $2\times10^3$ independent dynamical realizations in a fixed double-layer network and average on $30$ different double-layer networks to obtain these final densities of each state.

In Ref.~\cite{Shu:2015}, the variability measure has been verified to be very effective in identifying the SIR epidemic thresholds on various networks. However, for the interacting spreading dynamics, the interplay between them introduces a large external fluctuation into the respective spreading dynamics~\cite{Argollo:2004}, thus invalidate the variability measure. Therefore, we only qualitatively analyze the impact of the value of $\alpha$ (depicting the social reinforcement effect) on the outbreaks of information and disease. In the following simulations, we respectively define the reference information threshold ($\lambda_{Ae}$) and the reference epidemic threshold ($\lambda_{Be}$) to valuate the outbreak possibility. At the reference threshold, the outbreak rate just reaches a reference value (e.g., 0.01 or 0.05) by using a tolerance~\cite{Ferreri:2014}. The larger the value of reference information (epidemic) threshold, the harder the outbreak of the information (epidemic).

\begin{figure}
\begin{center}
\includegraphics[width=6.5in]{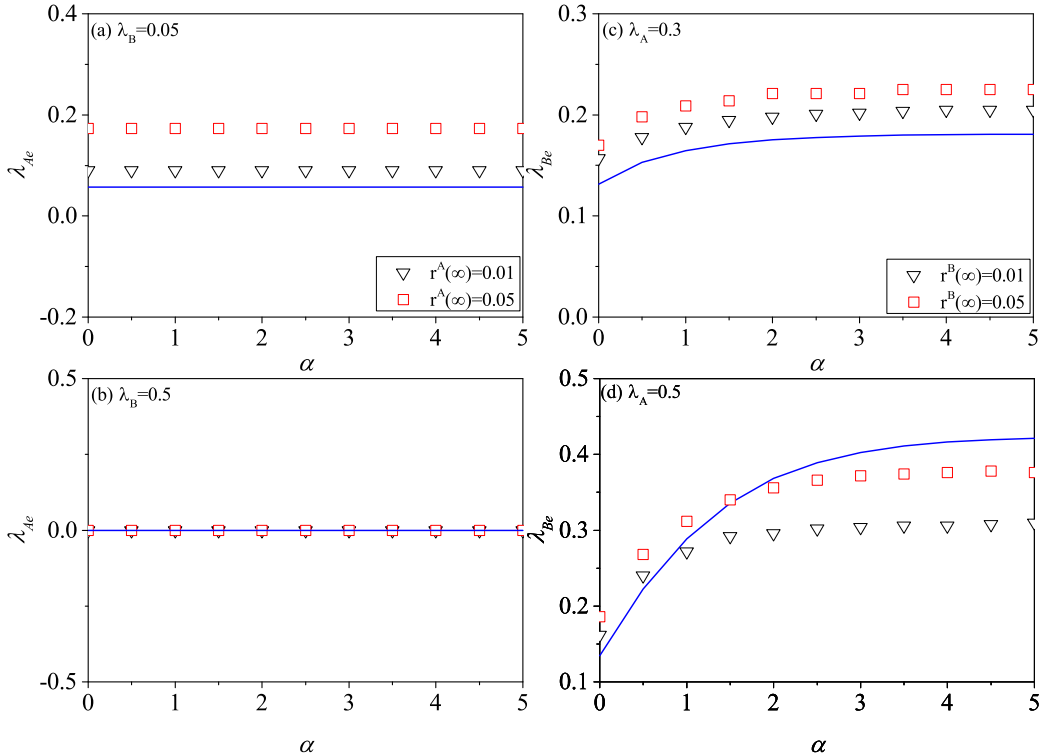}
\caption{\textbf{The impacts of social reinforcement effect on the outbreak threshold.} For SF-ER double-layer networks, the reference information threshold $\lambda_{Ae}$ and the reference epidemic threshold $\lambda_{Be}$ as the function of the value of $\alpha$ are obtained by numerical simulations. Owing to the difficulty of determining the threshold values from numerical predictions, the reference density, for which the final recovery density in layer $A$ ($B$) are 0.01 (black down triangles) and 0.05 (red squares), are set to be the reference threshold values. The blue solid line is the corresponding theoretical prediction from Eqs. (11)-(13) and (15). (a) In communication layer $A$, the reference information threshold $\lambda_{Ae}$ performs as a function of $\alpha$ for $\lambda_B=0.05$; (b) In communication layer $A$, the reference information threshold $\lambda_{Ae}$ performs as a function of $\alpha$ at $\lambda_B=0.5$; (c) In the physical contact layer $B$, the reference epidemic threshold $\lambda_{Be}$ performs as a function of $\alpha$ for $\lambda_A=0.3$; (d) In the physical contact layer $B$, the reference epidemic threshold $\lambda_{Be}$ performs as a function of $\alpha$ at $\lambda_A=0.5$.}
\label{fig2}
\end{center}
\end{figure}

From Figs.~\ref{fig2}(a) and (b), it can be seen that the impacts of the value of $\alpha$ on the reference information threshold $\lambda_{Ae}$ in layer $A$ can almost be ignored. Nevertheless, it is shown by Figs.~\ref{fig2}(c) and (d) that $\alpha$ has a remarkable influence on the reference epidemic threshold $\lambda_{Be}$ in layer $B$ when the information spreads faster than the disease. In particular, the epidemic threshold first increases with the value of $\alpha$, but then tends to be stable when the value of $\alpha$ increases. The greater value of $\alpha$ leads to the stronger reinforcement effect (i. e., the greater adoption probability $\xi_M$) in layer $A$, which thus can more effectively suppress the outbreak of epidemic in layer $B$. However, with the increasing of $\alpha$, the reinforcement effect of multiple information will reach a saturation point due to the restriction of network structure (e.g., mean degree and degree distribution) and information diffusion (e.g., transmission rate and recovery rate). Comparing Fig.~\ref{fig2}(d) with Fig.~\ref{fig2}(c), it can be seen that a larger value of $\lambda_A$ also causes a higher reference epidemic threshold $\lambda_{Be}$ (i.e., the disease transmission probability at which the final infection density reaches a fixed value such as $r^B(\infty)=0.01,0.05$).

\begin{figure}
\begin{center}
\includegraphics[width=6in]{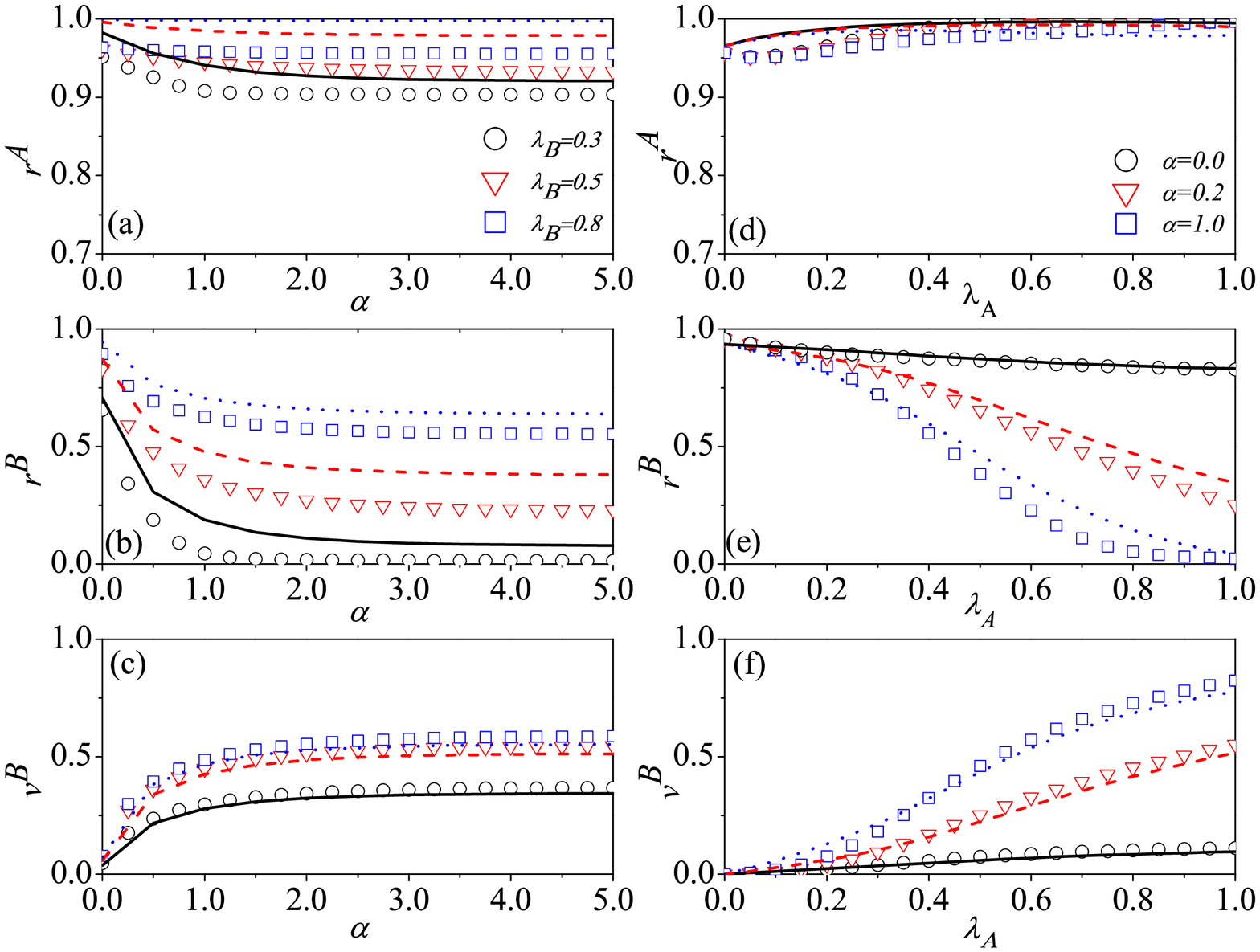}
\caption{\textbf{The impacts of social reinforcement effect and information transmission rate on
final states.} For SF-ER double-layer network, subfigures (a), (b), and (c) show the values of $r^A$, $r^B$ and $v^B$ as a function of $\alpha$ with different values of $\lambda_B$ (0.3, 0.5, and 0.8), and their analytical predictions are corresponded to the black solid, red dashed, and blue doted lines, respectively. Where $\lambda_A$ is set as $0.5$. Subfigures (d), (e), and (f) illustrate the values of $r^A$, $r^B$ and $v^B$ versus the parameter $\lambda_A$ for different values of $\alpha$ (0, 0.2, and 1.0), corresponding to the black solid, red dashed, and blue doted lines respectively. When $\lambda_B$ is fixed at $0.5$.}
\label{fig3}
\end{center}
\end{figure}

It is shown by Figs.~\ref{fig3}(a)-(c) that with different values of $\lambda_B$, more nodes in layer $B$ will be vaccinated [see Fig.~\ref{fig3}(c)] with the increase of parameter $\alpha$, leading to the spreading of epidemic in layer $B$ to be reduced or eliminated [see Fig.~\ref{fig3}(b)]. Moreover, the reduction of epidemic also decreases the number of informed individuals [see Fig.~\ref{fig3}(a)], i.e., $r^A$ is reduced too. It can also be seen from Figs.~\ref{fig3}(a)-(c) that $\alpha$ has a big influence on the values of $r^A$, $r^B$ and $v^B$ when $\alpha\in(0,1)$, but little influence when $\alpha\in[1,5]$. Figs.~\ref{fig3}(d)-(f) demonstrate the effects of $\lambda_A$ on $r^A$, $r^B$ and $v^B$ with different values of $\alpha$. From Fig.~\ref{fig3}(d), it can be found that $r^A$ decreases with $\lambda_A$ when $\lambda_A$ increases from zero, which is somewhat non-intuitive. As we know, when $\lambda_A$ increases from zero, the spreading of information quickly inhibits the spreading of epidemic, which also reduces the promotion effect of epidemic on information spreading. Moreover, the competing effects of the two aspects (the enhancement of information spreading due to the increase of $\lambda_A$ and the drop of information spreading due to the reduction of epidemic) may lead to the reduction of $r^A$. However, as we further increase the value of $\lambda_A$, the information can spread quickly and more individuals will be informed [see Fig.~\ref{fig3}(d)], which induces more people to be correspondingly vaccinated [see Fig.~\ref{fig3}(f)], naturally, the number of infected individuals is reduced [see Fig.~\ref{fig3}(e)]. It is noted that there are some discrepancies between the theoretical predictions and simulation results in Fig.~\ref{fig3}, because the developed mean field theory can't accurately capture the dynamical correlations between the two layers~\cite{Wei:2014}.

\begin{figure}
\begin{center}
\includegraphics[width=5in]{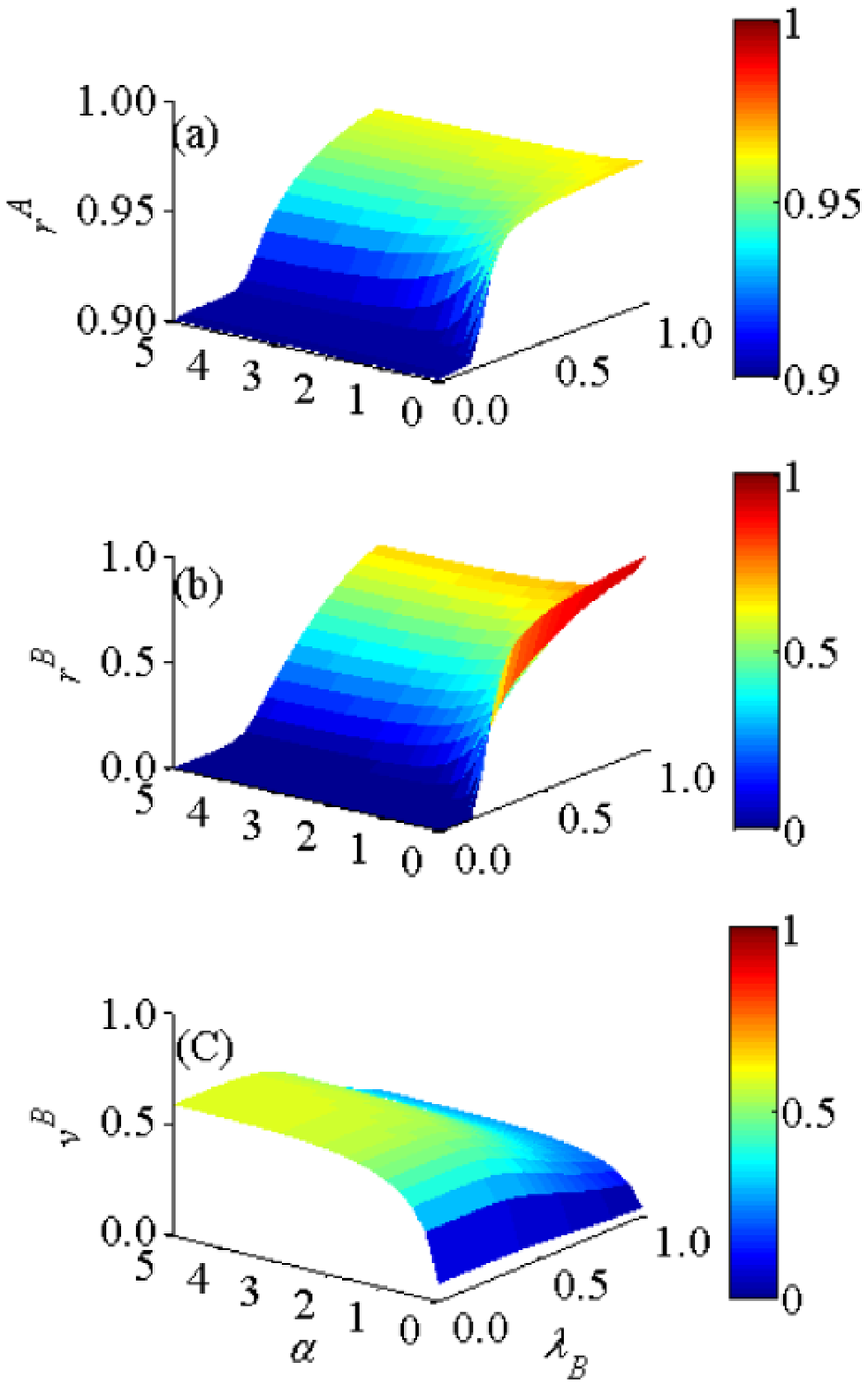}
\caption{\textbf{A systematic investigation of the impacts of social reinforcement effect and disease transmission rate on final states.} For SF-ER double-layer network, (a) recovered density $r^A$,
(b) recovered density $r^B$, (c) the vaccination density $v^B$ versus $\alpha$ and $\beta_B$ for $\lambda_A=0.5$.}
\label{fig4}
\end{center}
\end{figure}

We then further study the effects of $\alpha$ and $\lambda_B$ on the values of $r^A$, $r^B$ and $v^B$ in Fig.~\ref{fig4}. From Figs.~\ref{fig4} (a) and (b), it can be seen that, though the values of $r^A$ and $r^B$ increase with $\lambda_B$ as $\lambda_B>\lambda_{Bu}$, their growth rate slows down with larger $\alpha$. Fig.~\ref{fig4} demonstrates that increasing $\alpha$ can stimulate more individuals to take vaccination, thus raising the value of $v^B$. In RR-ER and SF-SF double-layer networks, the impact of social reinforcement effect on asymmetric interacting spreading dynamics is also explored and the obtained conclusion is consistent (see Figs. S1-S3 and Figs. S5-S7 in Supporting Information).

\begin{figure}
\begin{center}
\includegraphics[width=6in]{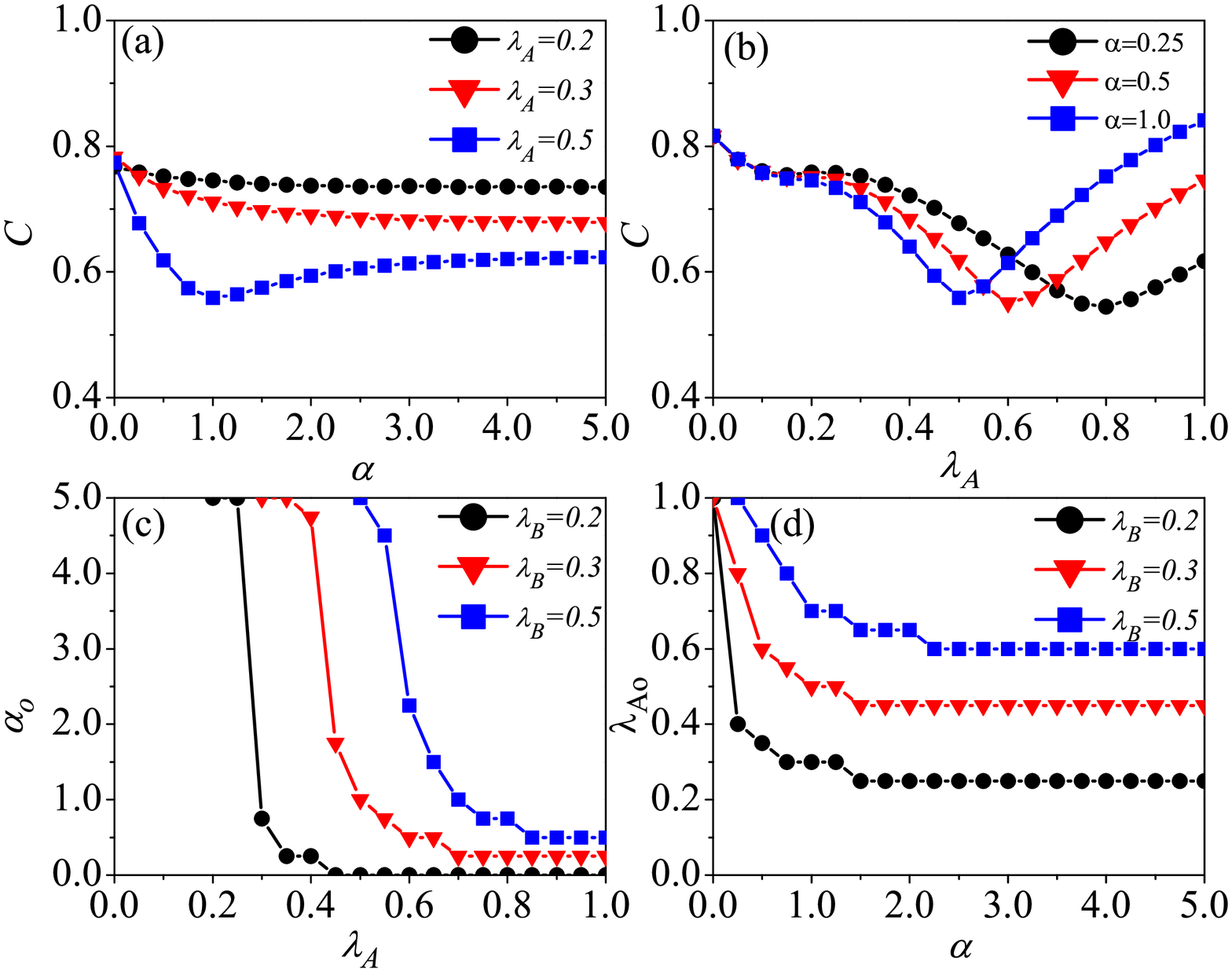}
\caption{\textbf{Impacts of social reinforcement effect and information transmission rate on the social cost and optimal control.} For SF-ER double-layer network, the social cost $C$ is versus the parameters of $\alpha$ and $\lambda_A$ in subfigures (a) and (b), respectively. Here, the value of $\lambda_B$ is fixed at $0.3$. The optimal $\alpha_o$ versus $\beta_A$ and optimal $\lambda_{Ao}$ versus $\alpha$ are demonstrated in subfigures (c) and (d), respectively. In (a), three different values of $\lambda_A$(0.2, 0.3, and 0.5) are selected, corresponding to the black circle solid, red triangle solid, and blue square solid lines, respectively. In (b), different values of $\alpha$ (0.25, 0.5 and 1.0) corresponds to the black circle solid, red triangle solid, and blue square solid lines, respectively. (c) the $\alpha_o$ versus $\lambda_A$ and (d) the $\lambda_{Ao}$ versus $\alpha$ under different $\lambda_B$ (0.2, 0.3 and 0.5)  correspond to the black circle solid, red triangle solid, and blue square solid lines, respectively.}
\label{fig5}
\end{center}
\end{figure}

\textbf{Social cost}. Measures to prevent or eliminate diseases~\cite{Stohr:2004,Reluga:2006,Mbah:2012} often mean certain social cost~\cite{Altarelli:2014,ZhangHF:2014}, such as, the cost of treating infected individuals and vaccinating susceptible individuals, cost of isolation, cost of reducing outgoing and so on. Although the rapid spread of information and the strong social reinforcement effect can effectively promote the vaccination behavior and thus suppress epidemic spreading, the total cost of vaccination will be greatly increased. From an overall perspective, the government wants to control the diseases to the greatest extent with the minimal cost. In doing so, we define the social cost~\cite{Altarelli:2014} as
\begin{eqnarray}
\label{A31}
C=\frac{\sum_{i\in\Lambda}(V_{B,i}c_V+R_{B,i}c_R)}{N},
\end{eqnarray}
here, $\Lambda$ is the set of all nodes in layer $B$. $V_{B,i}=1$ denotes the node $i$ is in $V$ state, otherwise, $V_{B,i}=0$. In the same way, $R_{B,i}=1$ means node $i$ has recovered from disease, otherwise, $R_{B,i}=0$. Since every node in layer $B$ can be in one of the three states: susceptible, recovered or vaccinated, it is impossible for $V_{B,i}$ and $R_{B,i}$ equaling to one at the same time. $c_V$ or $c_R$ denotes the cost of vaccination or treatment for a node. For the sake of simplicity, we assume the cost of vaccination and the cost of treatment are comparable and set both of them as unit for all individuals~\cite{Dybies:2004,Kleczkowski:2006}, \emph{i.e.,} $c_V=c_R=1$, and in this case, $C=r^B+v^B$.

Now we want to know how social reinforcement effect and information diffusion affect the social cost. Figs.~\ref{fig5}(a) and~\ref{fig5}(b) present the social cost $C$ as a function of the sensitivity factor $\alpha$ and the effective information transmission rate $\lambda_A$, respectively. As shown in Fig.~\ref{fig5}(a), there exists an optimal value of $\alpha$ which can guarantee the minimal social cost when $\lambda_A$ is larger than $\lambda_B$ (e.g., $\lambda_A=0.5$). However, with $\lambda_A<\lambda_B$, increasing $\alpha$ can reduce the social cost to some extent because the epidemic spreading is suppressed more or less. Also, there exists an optimal value of $\lambda_A$ leading to the minimal social cost [see Fig.~\ref{fig5}(b)]. When the number of vaccinated nodes is few, each vaccinated node can protect more than one node from infection, \emph{i.e.}, the herd immunity effect can be successfully produced when $V_B$ is small. Thus, increasing the value of $\alpha$ or $\lambda_A$ stimulates more vaccinated nodes, which can effectively reduce the social cost. With further increasing the number of vaccinated nodes the disease can be controlled to a very low level. Apparently, it is unnecessary to increase the vaccination coverage any more, because the total social cost will be increased again when $V_B$ is further increased. Therefore, an optimal vaccination coverage (i.e., optimal values of $\alpha$ and $\lambda_A$) can be gained by employing the two competing effects, thus guaranteeing the minimal social cost. Consistent conclusions are also obtained in analyzing the influence of social reinforcement effect and information diffusion on social cost in RR-ER double-layer and SF-SF double-layer networks (see Fig. S4 and Fig. S8 in Supporting Information). This suggests that reasonably control the social reinforcement effect and the spread of information is very critical to minimizing the total social cost. For the social reinforcement effect, the risk of disease cannot be ignored, neither should it be exaggerated. As to the spread of disease information, the government should not only ensure the rapid spread of it but also avoid the excessive spread of it. In Fig.~\ref{fig5}(c) [(d)], with the increase of $\lambda_A$ ($\alpha$), the optimal $\alpha_o$ ($\lambda_{Ao}$) is reduced, which means that with a faster spread of information (a stronger social reinforcement effect), a minimal social cost is required for a weaker social reinforcement effect (a slower spread of information).

Usually, different relative costs of vaccination and treatment are required for different diseases~\cite{Altarelli:2014,Meltzer:1999,Weycker:2004}. Considering the self-interest characteristic of individuals in real society~\cite{Galvani:2007}, the behavior of taking vaccination is unnecessary for individuals if the cost of vaccination surpasses that of treatment. Therefore, the cost of treatment is considered to be greater than that of vaccination~\cite{Dybies:2004, Kleczkowski:2012}. The impacts of different relative costs of vaccination and treatment (e.g., $c_R/c_V$=2 in Fig. S9 and $c_R/c_V$=5 in Fig. S10) on the optimal control are also studied in Supporting Information. It is found that the above conclusion remains unchanged qualitatively, but further study is still required~\cite{Kleczkowski:2012}.

\section*{Discussion}
In summarize, in this paper, a memory-based complex adoption mechanism was introduced into an asymmetrically interacting, double-layer network model to elucidate the mutual effects among information diffusion, epidemic spreading and the complex vaccination adoption mechanism. In the model, the information propagation and epidemic spreading occur in layer $A$ and layer $B$, respectively. Moreover, the probability of vaccination for each informed individual depends on the times of information who has received and the social reinforcement effect.  A mean-field based analysis was developed to reveal the two intricate spreading dynamics and to verify results of extensive simulations. Our findings show that such a complex vaccination adoption behavior with non-markov characteristics can inhibit the spread of disease and increase the epidemic threshold in the contact layer. Furthermore, when we consider the cost of vaccination and cost of the treatment for infected individuals, we found that there exists an optimal memory reinforcement effect and an optimal transmission rate of information which can minimize the social cost.

The challenges of studying the intricate interplay between social and biological contagions in human populations are generating interesting science~\cite{Bauch:2013}. In this work, we just considered the social reinforcement effect of cumulative information in complex adoption behavior and thus studied its impact on the two interacting spreading dynamics. As a matter of fact, the behavioral response to disease is also affected by socioeconomic factors such as psychological reflection, economic cost and infection status. The adoption behavior thus presents a more complex and diverse response mode, which may remarkably influence the asymmetric interacting spreading dynamics, especially for epidemic spreading. Our efforts along this line would stimulate further studies in the more realistic situation of asymmetric interactions.

\section*{Methods}

\textbf{Mean-Field equations for the spreading dynamics in layer $A$}. To derive the mean-field rate equations for the density variables, we considered the probabilities that $S_A$ ($S_B$) node is informed (infected) during the small time interval $[t,t+dt]$. According to the description of information spreading processes in two interacting spreading dynamical processes, it can be known that the loss of $s_{k_A}^A(t)$ (i.e., the density of the susceptible nodes with degree $k_A$ ) is caused by two aspects:
1) a $S_A$ node has received one or more pieces of information from its neighbors in layer $A$,\emph{ i.e.}, the node is informed by its neighbors;
2) a $S_A$ node's counterpart node in layer $B$ is susceptible (\emph{i.e.}, $S_B$), and it is infected by the disease at this time step.

In random configuration networks without degree correlations, for a $S_A$ node, the probability that one randomly selected neighbor is in $I_A$ state~\cite{Newman:2010} is given as
\begin{eqnarray}
\label{A17}
{\Theta_{S_A}^A(t)} & = & \frac{\sum_{k_A^{\prime}}(k_A^{\prime}-1)P_A(k_A^{\prime})\rho_{k_A^{\prime}}^A(t)}{\langle {k_A}\rangle},
\end{eqnarray}
where
\begin{eqnarray}
\label{A18}
{\rho_{k_A^{\prime}}^A(t)} & = & \sum_m{\rho_{k_A^{\prime}}^A(m,t)}
\end{eqnarray}
is the density of $I_A$ nodes with degree $k'_A$ at time $t$, and $\rho_{k_A^{\prime}}^A(m,t)$ is the density of $I_A$ nodes with degree $k'_A$ which have received $m$ pieces of information till time $t$. One should note that, $k_A'-1$ was adopted rather than $k_A'$ in Eq.~(\ref{A17}). For a $S_A$ node, since all of its neighbors cannot be informed by the $S_A$ node, one of its infected neighbors with degree $k_A'$ concedes a possibility that other $k_A'-1$ links connect to the $S_A$ node, excluding the link between this infected neighbor and its parent infected node. If we neglect the dynamical correlations between neighborhood, for a $S_A$ node, the probability for the node to have $n$ number of $I_A$ neighbors is
\begin{eqnarray}
\label{A19}
{\pi_{S_A,k_A}^A(n)} & = & B_{k_A,n}[\Theta_{S_A}^A(t)],
\end{eqnarray}
where $B_{k,m}(q)$ denotes the binomial factor $\binom{k}{m}{q}^m{(1-q)}^{k-m}$ . Based on the above factors, the probability of a $S_A$ node with degree $k_A$ to receive the information at least once is
\begin{eqnarray}
\label{A20}
{\Psi_{S_A,k_A}^A}(t) & = & \sum_{n=1}^{k_A}\pi_{S_A,k_A}^A(n)[1-{(1-\beta_A)}^n].
\end{eqnarray}

Similar to Eqs.~(\ref{A17}), (\ref{A18}) and (\ref{A19}), for a $S_B$ node in layer $B$, the probability that one randomly selected neighbor is in $I_B$ state is
\begin{eqnarray}
\label{A21}
{\Theta_{S_B}^B(t)} & = & \frac{\sum_{k_B^{\prime}}(k_B^{\prime}-1)P_B(k_B^{\prime})\rho_{k_B^{\prime}}^B(t)}{\langle {k_B}\rangle},
\end{eqnarray}
and
\begin{eqnarray}
\label{A22}
{\pi_{S_B,k_B}^B(n)} & = & B_{k_B,n}(\Theta_{S_B}^B(t))
\end{eqnarray}
is the probability of a $S_B$ node with degree $k_B$ which has $n$ number of $I_B$ nodes in his neighborhood. Moreover, the probability of the $S_B$ node with degree $k_B$ to be infected is
\begin{eqnarray}
\label{A23}
{\Psi_{S_B,k_B}^B}(t) & = & \sum_{n=1}^{k_B}\pi_{S_B,k_B}^B(n)[1-{(1-\beta_B)}^n].
\end{eqnarray}

At time step $t$, the density of $\rho_{k_A}^{A}(m,t)$ can be altered by two opposite cases: 1) for a node that is in $S_A$ state before time step $t$ and simultaneously receives $m$ pieces of information at time $t$ or that has received $n$ ($0<n<m$) pieces of information and simultaneously receives $m-n$ pieces of information at time $t$, leading to the gains of $\rho_{k_A}^{A}(m,t)$; 2) for an $I_A$ node
which has received $m$ pieces of information, and then receives one or more pieces of information again at time $t$, or the $I_A$ node recoveries to $R_A$ node, resulting in the losses of $\rho_{k_A}^{A}(m,t)$.

For an $I_A$ node, the probability that one selected neighbor is in $I_A$ state is given as
\begin{eqnarray}
\label{A24}
{\Theta_{I_A}^A(t)} & = & \frac{\sum_{k_A^{\prime}}k_A^{\prime}P_A(k_A^{\prime})\rho_{k_A^{\prime}}^A(t)}{\langle {k_A}\rangle}.
\end{eqnarray}
Thus, the probability of an $I_A$ node with degree $k_A$ to have
$n$ ($n\leq k_A$) number of informed neighbors is
\begin{eqnarray}
{\pi_{I_A,k_A}^A(n)}& = & B_{k_A,n}[\Theta_{I_A}^A(t)].
\end{eqnarray}
As a result, the probability that the $I_A$ node has received at least one piece of information is
\begin{eqnarray}
\label{A13}
\Psi_{I_A,k_A}^A(t) & = & \sum_{n=1}^{k_A}\pi_{I_A,k_A}^A(n)[1-{(1-\beta_A)}^n].
\end{eqnarray}

\textbf{Mean-field equations for the spreading dynamics in layer $B$}. There are two cases which can lead to the decrease of $s_{k_B}^B(t)$, as follows: 1) a $S_B$ node is infected by its neighbors in layer $B$ with probability $\Psi_{S_B,k_B}^B(t)$; 2) The $S_B$ node goes to $V_B$ state because its counterpart node in layer $A$ is informed and is willing to take vaccination. Firstly, we can conclude that a node must be in $S_B$ state if its counterpart node in layer $A$ is in $S_A$ state. Ignoring the inter-layer degree correlations and dynamical correlations, the probability that the counterpart node of a node with degree $k_B$ has degree $k_A$ and is in $S_A$ state can be written as $P_A({k_A})s_{k_A}^A(t)$. Combining Eqs.~(\ref{A1}) and~(\ref{A19}), for a $S_A$ node of degree $k_A$ which has $n$ number of informed neighbors and has just received $q$ pieces of information at time $t$, the probability of taking vaccination is determined by the term $\pi_{S_A,k_A}^A(n)B_{n,q}({\beta_A})\xi_q$. Considering the different numbers of $n$ and $p$, the probability of an individual to adopt vaccination can be obtained as
\begin{eqnarray}
\label{A16}
\chi_{S_A,k_A}^A(t) & = & P_A({k_A})s_{k_A}^A(t)\sum_{n=1}^{k_A}\pi_{S_A,k_A}^A(n)\sum_{q=1}^nB_{n,q}(\beta_A)\xi_q.
\end{eqnarray}
Secondly, when a node of degree $k_B$ is in $S_B$ state with probability $s_{k_B}^B(t)$ in layer $B$, its counterpart node may have already been informed of $m$ pieces of information with probability $P_A({k_A})\rho_{k_A}^A(m,t)$, if inter-layer degree correlations and dynamical correlations are ignored. Accumulating different cases of $\rho_{k_A}^A(m,t)$, the probability of an individual to take vaccination can be given as
\begin{eqnarray}
\chi_{I_A,k_A}^A(t) & = & P_A({k_A})\sum_m\rho_{k_A}^A(m,t)\sum_{n=1}^{k_A}\pi_{I_A,k_A}^A(n)\sum_{q=1}^nB_{n,q}(\beta_A)\xi_{m+q}.
\end{eqnarray}

\section*{Figure legends}

\textbf{Figure 1}:\textbf{Illustration of asymmetrically coupled spreading processes in a double-layered communication-contact network.} (a) Communication and contact networks, denoted respectively as layer $A$ and layer $B$, each have four nodes. Each node of layer $A$ is matched one-to-one with that of layer $B$ randomly. A node $i$ in layer $A$ is represented as ${A_i^M}$, where the subscript and superscript respectively represent the index of node and the times of received information. (b) At $t=0$, node $B_1$ in layer $B$ is randomly chosen as the initial infected node and its counterpart, node $A_1$ in layer $A$, gains the information and becomes informed state and updates $M=1$. While all other pairs of nodes, one from layer $A$ and another from layer $B$, are in the susceptible state. (c) At $t=1$, node $B_3$ in layer $B$ can be infected by infected neighbor $B_1$ with probability $\beta_B$, and if it is indeed infected, its corresponding node $A_3$ in layer $A$ will get the information as well and update $M=1$. Within layer $A$ the information is transmitted from $A_1$ to $A_2$, with $M=1$ for $A_2$. Since, by this time, $A_2$ is already aware of the infection spreading, whereas its counterpart $B_2$ in layer $B$ takes vaccination with probability $\xi_1$, but fails. At the same time, node $A_1$ in layer $A$ and its counterpart $B_1$ in layer $B$ enter into the refractory state with probability $\mu_A$ and $\mu_B$, respectively. (d) At $t=2$, in layer $A$, $A_3$ successfully transmits the information to $A_2$. In this case, node $A_2$ updates $M=2$. At the same time, its counterpart $B_2$ in layer $B$ takes vaccination with probability $\xi_2$ and successfully becomes a vaccinated node. The spreading dynamics terminate as all infected/informed nodes have entered into the refractory state.

\textbf{Figure 2}:\textbf{The impacts of social reinforcement effect on the outbreak threshold.} For SF-ER double-layer networks, the reference information threshold $\lambda_{Ae}$ and the reference epidemic threshold $\lambda_{Be}$ as the function of the value of $\alpha$ are obtained by numerical simulations. Owing to the difficulty of determining the threshold values from numerical predictions, the reference density, for which the final recovery density in layer $A$ ($B$) are 0.01 (black down triangles) and 0.05 (red squares), are set to be the reference threshold values. The blue solid line is the corresponding theoretical prediction from Eqs. (11)-(13) and (15). (a) In communication layer $A$, the reference information threshold $\lambda_{Ae}$ performs as a function of $\alpha$ for $\lambda_B=0.05$; (b) In communication layer $A$, the reference information threshold $\lambda_{Ae}$ performs as a function of $\alpha$ at $\lambda_B=0.5$; (c) In the physical contact layer $B$, the reference epidemic threshold $\lambda_{Be}$ performs as a function of $\alpha$ for $\lambda_A=0.3$; (d) In the physical contact layer $B$, the reference epidemic threshold $\lambda_{Be}$ performs as a function of $\alpha$ at $\lambda_A=0.5$.

\textbf{Figure 3}:
\textbf{The impacts of social reinforcement effect and information transmission rate on
final states.} For SF-ER double-layer network, subfigures (a), (b), and (c) show the values of $r^A$, $r^B$ and $v^B$ as a function of $\alpha$ with different values of $\lambda_B$ (0.3, 0.5, and 0.8), and their analytical predictions are corresponded to the black solid, red dashed, and blue doted lines, respectively. Where $\lambda_A$ is set as $0.5$. Subfigures (d), (e), and (f) illustrate the values of $r^A$, $r^B$ and $v^B$ versus the parameter $\lambda_A$ for different values of $\alpha$ (0, 0.2, and 1.0), corresponding to the black solid, red dashed, and blue doted lines respectively. When $\lambda_B$ is fixed at $0.5$.

\textbf{Figure 4}:
\textbf{A systematic investigation of the impacts of social reinforcement effect and disease transmission rate on final states.} For SF-ER double-layer network, (a) recovered density $r^A$,
(b) recovered density $r^B$, (c) the vaccination density $v^B$ versus $\alpha$ and $\beta_B$ for $\lambda_A=0.5$.

\textbf{Figure 5}:
\textbf{Impacts of social reinforcement effect and information transmission rate on the social cost and optimal control.} For SF-ER double-layer network, the social cost $C$ is versus the parameters of $\alpha$ and $\lambda_A$ in subfigures (a) and (b), respectively. Here, the value of $\lambda_B$ is fixed at $0.3$. The optimal $\alpha_o$ versus $\beta_A$ and optimal $\lambda_{Ao}$ versus $\alpha$ are demonstrated in subfigures (c) and (d), respectively. In (a), three different values of $\lambda_A$(0.2, 0.3, and 0.5) are selected, corresponding to the black circle solid, red triangle solid, and blue square solid lines, respectively. In (b), different values of $\alpha$ (0.25, 0.5 and 1.0) corresponds to the black circle solid, red triangle solid, and blue square solid lines, respectively. (c) the $\alpha_o$ versus $\lambda_A$ and (d) the $\lambda_{Ao}$ versus $\alpha$ under different $\lambda_B$ (0.2, 0.3 and 0.5)  correspond to the black circle solid, red triangle solid, and blue square solid lines, respectively.

\section*{Acknowledgments}
This work was supported by the National Natural Science Foundation
of China (Grant Nos. 11105025, 11575041 and 61473001).

\section*{Author contributions}
Q. H. L., W. W., M. T. designed the experiments.
Q. H. L., W. W., M. T. and H. F. Z. analyzed the results.
Q. H. L., M. T. and H. F. Z. wrote the paper.

\section*{Additional information}



{\bf Competing financial interests}:
The authors declare no competing financial interests.

\newpage

\begin{center}
{\Large
Supporting Information for\\
\vspace{0.5cm}
\textbf{Impacts of complex behavioral responses on asymmetric interacting spreading dynamics in multiplex networks}
}\\
\vspace{0.5cm}

\large{Quan-Hui Liu, Wei Wang, Ming Tang and Hai-Feng Zhang}

\end{center}

\noindent${\Large\textbf{S1. Theoretical analysis}}$\\

The heterogeneous mean-field theory~\cite{sBarthelemy11:2004} was adopted to derive the mean-field equations for the uncorrelated double-layer network. Let $P_A(k_A)$ [$P_B(k_B)$] be the degree distribution of communication layer $A$ (contact layer $B$), and the average degrees of $A$ and $B$ are $\langle k_A\rangle=\sum_{k_A}k_AP_A(k_A)$ and $\langle k_B\rangle=\sum_{k_B}k_BP_B(k_B)$, respectively. Meanwhile, we assume that inner-layer links and inter-layer links have no degree correlations. The variables of $s_{k_A}^{A}(t)$, $\rho_{k_A}^{A}(t)$ and $r_{k_A}^{A}(t)$ are used to denote the densities of the susceptible, informed, and recovered nodes with degree $k_A$ in layer $A$ at time $t$, respectively. Thereinto, $\rho_{k_A}^{A}(t)=\sum_{m}\rho_{k_A}^{A}(m,t)$, and  $\rho_{k_A}^{A}(m,t)$ is the density of $I_A$ nodes with degree $k_A$ who has received $m$ pieces of information till time $t$. Similarly, $s_{k_B}^{B}(t)$, $\rho_{k_B}^{B}(t)$, $r_{k_B}^{B}(t)$ and $v_{k_B}^{B}(t)$ are the densities of the susceptible, infected, recovered and vaccinated nodes with degree $k_B$ in layer $B$ at time $t$, respectively.\\

\noindent${\Large\textbf{A. Mean-field rate equations}}$\\

The mean-field rate equation of the information spreading in layer $A$ is\\
$$\frac{ds_{k_A}^A(t)}{dt} = -s_{k_A}^A(t)[{\Psi_{S_A,k_A}^A}(t)+\sum_{k_B}P_B({k_B})\Psi_{S_B,k_B}^B(t)], \eqno (\text{S}1)$$
For $m=1$, the rate equation of $\rho_{k_A}^{A}(1,t)$ is given as\\
$$\frac{d\rho_{k_A}^A(1,t)}{dt}=s_{k_A}^A(t)\sum_{n=1}^{k_A}\pi_{S_A,k_A}^A(n)B_{n,1}(\beta_A)\\
+s_{k_A}^A(t)\sum_{k_B}P_B({k_B})\Psi_{S_B,k_B}^B(t)-\rho_{k_A}^A(1,t)\Psi_{I_A,k_A}^A(t)-\mu\rho_{k_A}^A(1,t),
\eqno (\text{S}2)$$\\
When $m>1$, the rate equation of $\rho_{k_A}^A(m,t)$ is described as\\
$$\frac{d\rho_{k_A}^A(m,t)}{dt} =  s_{k_A}^A(t)\sum_{n=m}^{k_A}\pi_{S_A,k_A}^A(n)B_{n,m}(\beta_A)\\
\nonumber  +  \sum_{q=1}^{m-1}\rho_{k_A}^A(q,t)\sum_{n=m-q}^{k_A}\pi_{I_A,k_A}^A(n)B_{n,m-q}(\beta_A)\\
\nonumber  - \rho_{k_A}^A(m,t)\Psi_{I_A,k_A}^A(t)-\mu\rho_{k_A}^A(m,t), \eqno (\text{S}3)$$\\
$$\frac{dr_{k_A}^A(t)}{dt}  =  \mu\sum_m\rho_{k_A}^A(m,t).\eqno (\text{S}4)$$\\

The mean-field rate equation of the epidemic spreading in layer $B$ is
$$\frac{ds_{k_B}^B(t)}{dt} =-{s_{k_B}^B(t)}\Psi_{S_B,k_B}^B(t)-\sum_{k_A}\chi_{S_A,k_A}^A(t)-s_{k_B}^B(t)\sum_{k_A}\chi_{I_A,k_A}^A(t),
\eqno (\text{S}5)$$
$$\frac{d\rho_{k_B}^B(t)}{dt} = s_{k_B}^B(t)\Psi_{S_B,k_B}^B(t)-\mu\rho_{k_B}^B(t), \eqno (\text{S}6)$$
$$\frac{r_{k_B}^B(t)}{dt} = \mu\rho_{k_B}^B(t), \eqno (\text{S}7)$$
$$\frac{dv_{k_B}^B(t)}{dt} = \sum_{k_A}\chi_{S_A,k_A}^A(t)+s_{k_B}^B(t)\sum_{k_A}\chi_{I_A,k_A}^A(t).\eqno (\text{S}8)$$

From Eqs. (S1)-(S8), the density associated with each distinct state in layer $A$ or $B$ is given by

$$x^H(t) = \sum_{k_H=1}^{k_H,max}P_H({k_H})x_{k_H}^H(t),$$
where $H\in\{A,B\}$, $x\in\{s,\rho,r,v\}$, and $k_{H,min}$ ($k_{H,max}$) denotes the smallest (largest) degree of layer $H$.
Specially, the density of $I_A$ node with degree $k_A$ in layer $A$ is $\rho_{k_A}^A(t)=\sum_{m}\rho_{k_A}^A(m,t)$.
The final densities of the whole system can be obtained by taking the limit $t\rightarrow\infty$.\\

\noindent${\Large\textbf{B. Linear analysis of information threshold in layer A}}$\\

On an uncorrelated nonoverlapping double-layer network, at the outset of the spreading dynamics, the whole system can be regarded
as consisting of two coupled SI-epidemic subsystems~\cite{sNewman:2010} with the time evolution described by  equations~(S2),(S3) and (S6). As $t\rightarrow 0$, one has $s_{k_A}^{A}(t)\approx 1$ and $s_{k_B}^{B}(t)\approx 1$, which reduce equations~(S2),(S3) and (S6)  as
$$
\begin{cases}
    \frac{d\rho_{k_A}^{A}(1,t)}{dt} =  \beta_{A}k_A\Theta_{S_A}^A(t) +
\beta_{B}\langle k_B\rangle\Theta_{S_B}^B(t)-\mu\rho_{k_A}^A(1,t), \\
\frac{d\rho_{k_A}^A(m,t)}{dt} = 0 ~( m>1 ), \\
\frac{d\rho_{k_B}^{B}(t)}{dt}  = \beta_{B}k_B\Theta_{S_B}^B(t)-\mu\rho_{k_B}^{B}(t).
\end{cases} \eqno (\text{S}9)
$$
The above equations can be simplified as matrix form:
$$\frac{d\vec{\rho}}{dt}=\frac{C\vec{\rho}}{\mu}-\vec{\rho}, \eqno (\text{S}10) $$
where
$$
\vec{\rho} \equiv (\rho_{k_A=1}^{A}(1),\ldots,\rho_{k_{A,max}}^{A}(1),\rho_{k_B=1}^{B},
\ldots,\rho_{k_{B,max}}^{B})^{T}.
\eqno (\text{S}11) $$
The matrix $C$ is written as a block matrix:
$$
C=\left(
\begin{array}{ccc}
    C^A & D^{B}\\
    0 & C^B\\
  \end{array}
\right),
\eqno(\text{S}12) $$
whose elements are given as
\begin{eqnarray}
\nonumber
C_{k_A,k'_A}^A & = & [\beta_A{k_A}({k'_A}-1)P_{A}(k'_A)]/{\langle k_{A}\rangle},\\
\nonumber
C_{k_B,k'_B}^B & = & [\beta_B{k_B}({k'_B}-1)P_{B}(k'_B)]/{\langle k_{B}\rangle},\\
\nonumber
D_{k_B,k'_B}^B & = & \beta_B({k'_B}-1)P_{B}(k'_B).
\end{eqnarray}
In general, information spreading in layer $A$ can be facilitated by the outbreak of the epidemic in layer $B$, since an infected node in layer $B$ instantaneously makes its counterpart node in layer $A$ ``infected'' by the information immediately and certainly. That is to say, the number of the informed nodes in layer $A$ is larger than the number of the infected nodes in layer $B$. If the maximum eigenvalue $\Lambda_{C}$ of matrix $C/\mu$ is greater than $1$, an outbreak of the information will occur absolutely~\cite{sSaumell:2012}. We then have

$$ \Lambda_C=\mbox{max}\{\Lambda_A, \Lambda_B\},\eqno(\text{S}13) $$
where max $\{\}$ denotes the greater of the two, and
\begin{eqnarray}
\nonumber
\Lambda_A  =  \beta_A(\langle {k_A}^2\rangle-\langle k_A\rangle)/(\mu\langle k_A\rangle),\\
\nonumber
\Lambda_B  =  \beta_B(\langle {k_B}^2\rangle-\langle k_B\rangle)/(\mu\langle k_B\rangle),
\end{eqnarray}
are the maximum eigenvalues of matrices $C_A$ and $C_B$~\cite{sMieghem:2011}, respectively. Thus, the outbreak threshold for the spreading in layer $A$ is given as
$$
\beta_{Ac} = \left\{\begin{array}{l}\beta_{Au},
\quad for \quad \beta_{B}\leq \beta_{Bu}; \\
0, \quad \quad for  \quad \beta_{B}> \beta_{Bu}.
\end{array} \right.
\eqno(\text{S}14) $$
Here $\beta_{Au}   \equiv   \mu\langle k_A\rangle/({\langle {k_A}^2\rangle
}-\langle k_A\rangle)$
and $\beta_{Bu}   \equiv
 \mu\langle k_B\rangle/(\langle {k_B}^2\rangle-\langle k_B\rangle)$
denote the outbreak threshold of information spreading in layer $A$ when it is isolated from layer $B$, and the outbreak threshold of epidemic spreading in layer $B$ when the coupling
between the two layers is absent, respectively.\\

\noindent${\Large\textbf{C. Competing percolation theory for epidemic threshold in layer $B$}}$\\

For $\beta_A<\beta_{Au}$, Eq. (S14) shows that the information cannot break out in layer $A$ if layer $A$ and layer $B$ are isolated. When the two spreading dynamics are interacting,\emph{ near the epidemic threshold}, the spread of epidemic in layer $B$ can only lead to a few of counterpart nodes in layer $A$ ``infected'' with the information, and thus these informed nodes in layer $A$ have negligible effect on the epidemic dynamics in layer $B$ since $\beta_A<\beta_{Au}$. The above explanation indicates that $\beta_{Bc}\approx\beta_{Bu}$ when $\beta_A<\beta_{Au}$. However, for $\beta_{A}>\beta_{Au}$, the information outbreak in layer $A$ which makes many counterpart nodes in layer $B$ vaccinated, thus hinders the spread of epidemic in layer $B$.
Once a node is in the vaccination state, it will no longer be infected. Usually, we can regard this kind of vaccination as a type of ``disease,'' and every node in layer $B$ can be in one of the two states: infected or vaccinated. Epidemic spreading and vaccination diffusion (derived by information diffusion) can thus be viewed as a pair of competing ``diseases'' spreading in layer $B$~\cite{Karrer:2011}. As pointed out by Karrer and Newman~\cite{Karrer:2011}, in the limit of large network size $N$ and the two competing diseases with different growth rates, then they can be treated as if they were in fact spreading non-concurrently, one after the other.

To clarify the interplay between epidemic and vaccination spreading, we should determine which one is the faster ``disease''. At the early stage, the average number of infected nodes in the isolated layer $B$ grows exponentially as $N_{e}(t)=n_0(R_{e})^{t}=n_0e^{t\ln R_{e}}$, where $R_{e}=\beta_{B}/\beta_{Bu}$ is the basic reproductive number for the disease in the isolated layer $B$~\cite{sNewman:2010}, and $n_0$ denotes the number of initially infected nodes. Similarly, for information spreading in the isolated layer $A$, the average number of informed nodes at the early time is $N_{i}(t)=n_1(R_{i})^{t}=n_1e^{t\ln(R_{i})}=N\sum_{m}\rho^{A}(m,t)$, where $n_0=n_1$, $\rho^A(m,t)=\sum_{k_A}P_A(k_A)\rho_{k_A}^A(m,t)$ denotes the density of the nodes who have received $m$ pieces of information till time step $t$, and $R_{i}=\beta_{A}/\beta_{Au}$ is the reproductive number for information spreading in the isolated layer $A$. So the number of vaccination nodes is $N_V(t)=N\sum_{m}\rho^A(m,t)\xi_m$, which is larger than $\xi_1n_0e^{t\ln(R_i)}$ since $\xi_m>\xi_1$, and which is smaller than $n_0e^{t\ln(R_i)}$ since $\xi_m<1$. As a result, at the early stage, we can view that $N_v$ grows exponentially and the growth satisfies $N_V\sim O(N_i)$.

Since the number of vaccination and infection both grow in an exponentially way, we can obtain the ratio of their growth rates as $$
\theta=\frac{R_i}{R_{e}}=\frac{\beta_A\beta_{Bu}}{\beta_B\beta_{Au}}. \eqno(\text{S}15) $$
When $\theta<1$, \emph{i.e.}, $\beta_B\beta_{Au}>\beta_A\beta_{Bu}$, the disease process grows faster than the vaccination process. In this case, the effect of vaccination is insignificant and can be neglected. However, when $\theta>1$, the information process spreads faster than the epidemic process, which is in accord with realistic situations since many on-line social networks and mass media can promote the spreading of information. Given that vaccination and epidemic can be treated successively and separately, by letting $\beta_B=0$ and obtaining the final density of vaccination $v^B(\infty)|_{\beta_B=0}$ from Eq.~({S8}), the threshold of epidemic outbreak is given as~\cite{sWei:2014}
$$
{\beta_{Bc}}=\frac{\mu\langle k_B\rangle}{[1-v^B(\infty)|_{\beta_B=0}]({\langle {k_B}^2\rangle}-\langle k_B \rangle)}.
\eqno(\text{S}16) $$

\noindent${\Large\textbf{S2. Simulation results}}$\\

We first describe the simulation processes of the two spreading dynamics in double-layer networks, and then present results for RR-ER double-layer and SF-SF double-layer networks. Lastly, we study the effect of different relative cost of vaccination and treatment on total social cost in SF-ER double layer networks.\\

\noindent${\Large\textbf{A. Simulation process}}$\\

To initiate an epidemic spreading process, a node in layer $B$ is randomly
infected and its counterpart node in layer $A$ is thus in the informed state,
too. The updating process is performed with parallel dynamics, which is widely
used in statistical physics~\cite{sMarro:1999}. At each time step, we first
calculate the informed (infected) probability $\pi_{A}=1-(1-\beta_A)^{n_I^A}$ [$\pi_B=1-(1-\beta_B)^{n_I^B}$] that each susceptible or informed node in layer $A$
may be informed or informed again by its informed neighbors and each susceptible node in
layer $B$ infected by its infected neighbors, where $n_I^A$ ($n_I^B$) is the number of its
informed (infected) neighboring nodes.

According to the dynamical mechanism, once node $A_i$ is in the susceptible state,
its counterpart node $B_i$ will be also in the susceptible state.
Besides, when a node in layer $A$ is in the informed state, its counterpart
node may be in the susceptible state. Considering the asymmetric coupling between
the two layers in these two cases, both the information-transmission and disease-transmission events
can hardly occur at the same time. Thus, with probability $\pi_A/(\pi_A+\pi_B)$,
node $A_i$ have a probability $\pi_A$ to get the information from its informed neighbors in layer $A$.
If node $A_i$ is informed, its counterpart node $B_i$
will turn into the vaccination state with probability $\xi_m$, where $m$ is total times of information the node has received.
With probability $\pi_B/(\pi_A+\pi_B)$, node $B_i$ have a probability $\pi_B$
to get the infection from its infected neighbors in layer $B$,
and then node $A_i$ also get the information about
the disease.

In the other case that node $B_i$ and its corresponding node $A_i$
are in the susceptible state and the informed (or refractory) state respectively,
only the disease-transmission event can occur at the time step.
Thus, node $B_i$ will be infected with probability $\pi_B$.

After renewing the states of susceptible nodes,
each informed (infected) node can enter the recovering phase with probability
$\mu=0.5$. The spreading dynamics terminates when all informed (or infected) nodes in both
layers are recovered, and the final densities $r^A$, $r^B$, and $v^B$ are
then recorded. The simulations are implemented using $30$ different double-layer network realizations and each realization is repeated $2\times10^3$ times.
The network size of $N_A=N_B=1 \times 10^4$ and average degrees
$\langle k_A\rangle=\langle k_B\rangle=8$ are used for all
subsequent numerical results, unless otherwise specified.\\

\noindent${\Large\textbf{B. RR-ER double-layer network}}$\\

In RR-ER double-layer network, we also investigate the impacts of social reinforcement effect
on the two types of spreading dynamics. At first, We use the standard configuration model~\cite{sNewman:2001-2} to generate regular random network (RR) for the communication subnetwork (layer A). The contact subnetwork in layer $B$ is of the Erd\H{o}s and R\'{e}nyi (ER) type~\cite{sErdos:1959}. We use the notation RR-ER to denote the double-layer network. The sizes of both layers are set to be $N_A=N_B=1 \times 10^4$ and their average degrees are
$\langle k_A\rangle=\langle k_B\rangle=8$. And we set $\xi_1=0.05$, $\mu=0.5$ in the following simulations. As shown in Figs. S1, S2, S3 and S4, we obtain the similar results of social reinforcement effect on the two types of spreading dynamics as in SF-ER double network.\\

\noindent${\Large\textbf{C. SF-SF double-layer network}}$\\

In SF-SF double-layer network, we also investigate the impacts of social reinforcement effect
on the two types of spreading dynamics. At first, We use the standard configuration model to
generate networks with power-law degree distributions~\cite{sNewman:2005-2,sNewman:2001-2,sCatanzaro:2005} for the communication subnetwork (layer A), with
$P_A(k_A)=\zeta{k_A^{-\gamma_A}}$, $\zeta=1/\sum_{k_{min}}^{k_{max}}{k_A^{-\gamma_A}}$,  $\gamma=3.0$ and the maximum degree $k_{max}{\sim}N^{1/(\gamma_A-1)}$. The contact subnetwork in layer $B$
is generated with the same methods as layer $A$. We
use the notation SF-SF to denote the double-layer network. The sizes of both
layers are set to be $N_A=N_B=1 \times 10^4$ and their average degrees are
$\langle k_A\rangle=\langle k_B\rangle=8$. And we set $\xi_1=0.05$, $\mu=0.5$ in the following simulations. As shown in Figs. S5, S6, S7 and S8, we obtain the similar results of social reinforcement effect on the two types of spreading dynamics as in SF-ER double network.\\

\noindent${\Large\textbf{D. Different relative cost of vaccination and treatment}}$\\

We study the different relative costs of vaccination and treatment to the effect of optimal control in SF-ER double-layer networks. we have assumed that the cost of treatment is twice and five times of vaccination cost, as shown in Fig. S9 and Fig. S10, respectively. We find when the information spreads faster than the disease, there still exists an optimal $\alpha$ yielding the least social cost. When the information about disease spreads slowly, increasing $\alpha$ can result in less social cost. These results have shown that the different relative costs of vaccination and treatment do not influence previous conclusion qualitatively.\\

\setcounter{figure}{0}
\begin{figure}
\epsfig{file=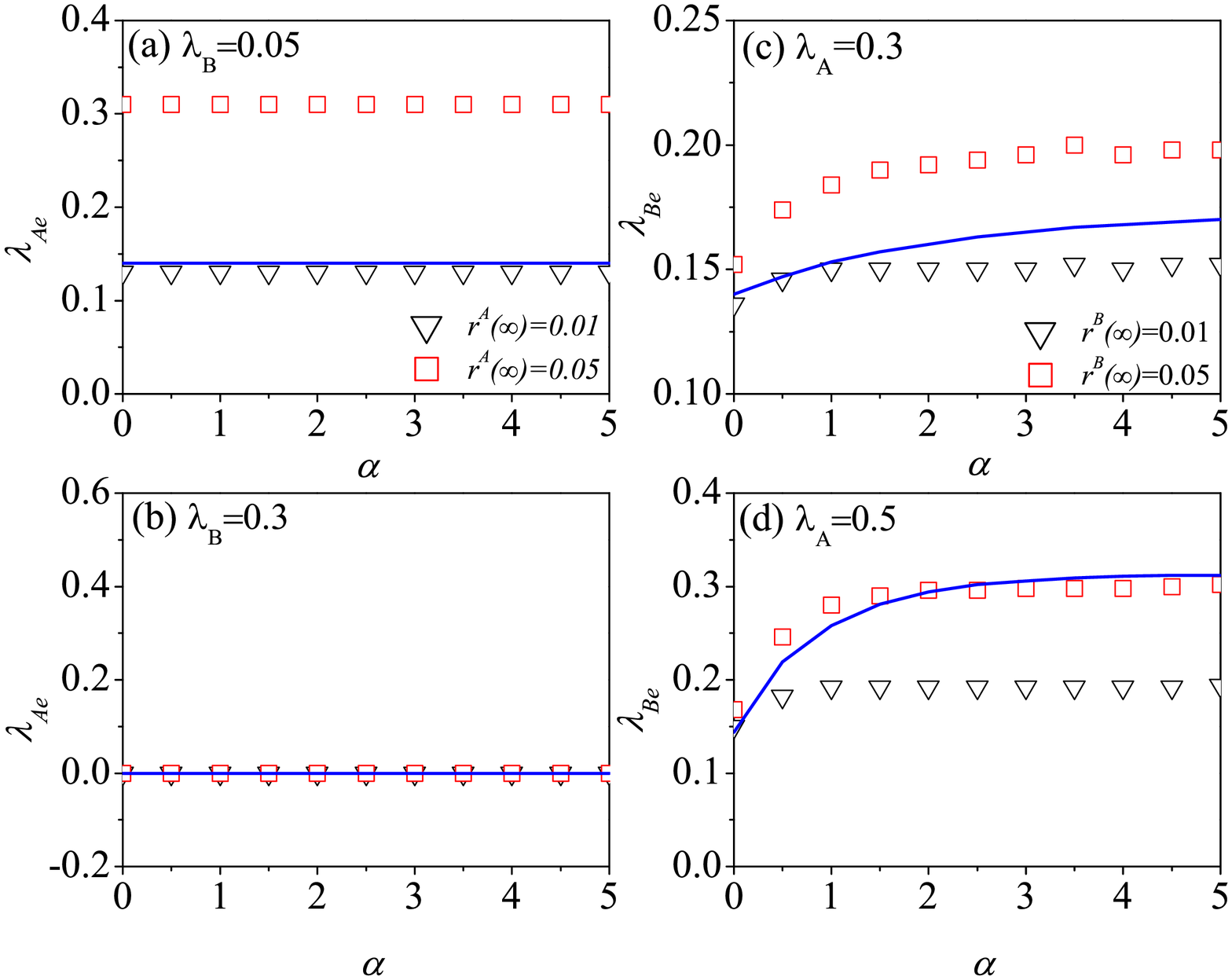,width=1\linewidth}
\renewcommand\thefigure{S\arabic{figure}}
\caption{\textbf{The impacts of social reinforcement effect on the outbreak threshold.} For RR-ER double-layer network, the reference information threshold $\lambda_{Ae}$ and the reference epidemic threshold $\lambda_{Be}$ as the function of $\alpha$ are obtained by numerical simulation. Owing to the difficulty of determining the threshold values from numerical predictions, we respectively take the critical density where the final recovery density in layer $A$ ($B$) are 0.01 (gray circles), 0.02 (oliver downtriangles) and 0.05(blue squares) as the reference threshold values. The red solid line is the corresponding theoretical prediction from Eqs. (S14) and (S16). (a) In communication layer $A$, the reference information threshold $\lambda_{Ae}$ as a function of $\alpha$ when $\lambda_B$ is set as $0.5$; (b) In physical contact layer $B$, the reference epidemic threshold $\lambda_{Be}$ as a function of $\alpha$ at $\lambda_A=0.5$.}
\label{figS1}
\end{figure}

\begin{figure}
\begin{center}
\epsfig{file=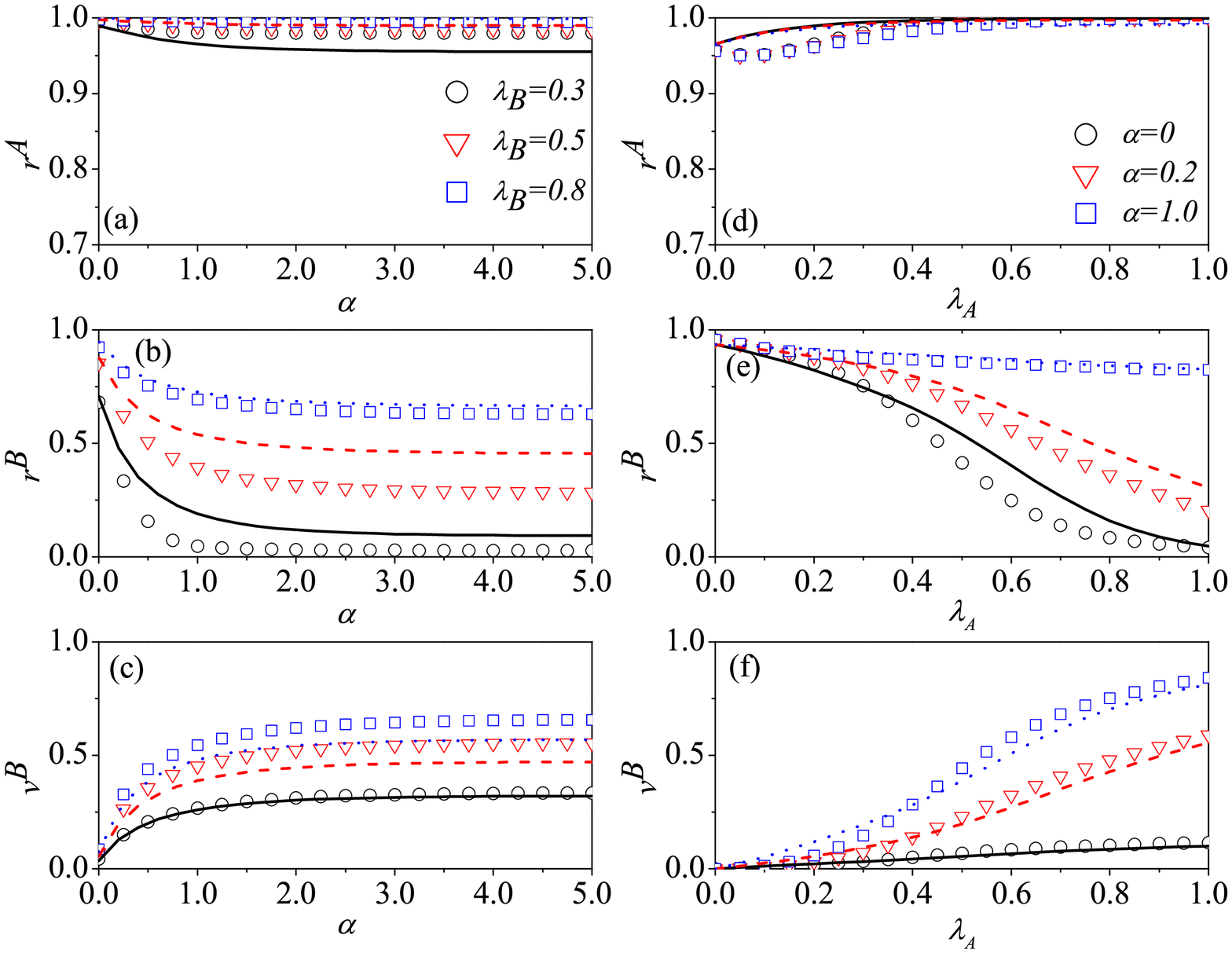,width=1\linewidth}
\renewcommand\thefigure{S\arabic{figure}}
\caption{\textbf{The impacts of social reinforcement effect and information transmission rate on final states.} For RR-ER double-layer network, subfigures (a), (b), and (c) show the values of $r^A$, $r^B$ and $v^B$ as a function of $\alpha$ for different values of $\lambda_B$ (0.3, 0.5, and 0.8), with the analytical predictions corresponding to the black solid, red dashed, and blue doted lines, respectively. When $\lambda_A$ is set as $0.5$. Subfigures (d), (e), and (f) illustrate the values of $r^A$, $r^B$ and $v^B$ versus the parameter $\lambda_A$ for different values of $\alpha$ (0, 0.2, and 1.0), corresponding to the black solid, red dashed, and blue doted lines respectively. When $\lambda_B$ is fixed at $0.5$.}
\label{figS2}
\end{center}
\end{figure}

\begin{figure}
\epsfig{file=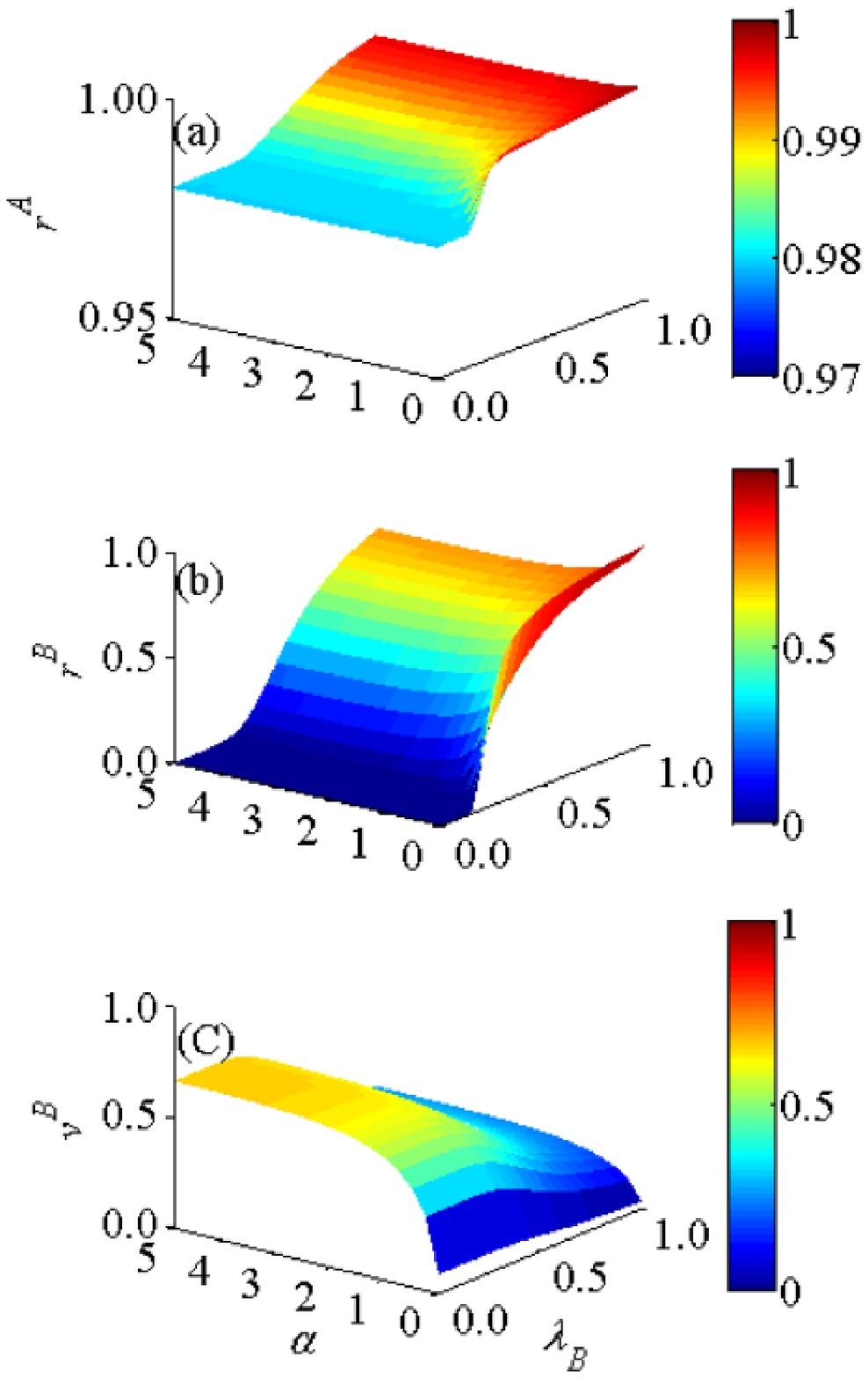,width=0.7\linewidth}
\renewcommand\thefigure{S\arabic{figure}}
\caption{\textbf{A systematic investigation of social reinforcement effect and disease transmission rate impact on final states.} For RR-ER double-layer network, (a) recovered density $r^A$,
(b) recovered density $r^B$, (c) the vaccination density $v^B$ versus $\alpha$ and $\beta_B$ for $\lambda_A=0.5$.}
\label{figS3}
\end{figure}

\begin{figure}
\begin{center}
\epsfig{file=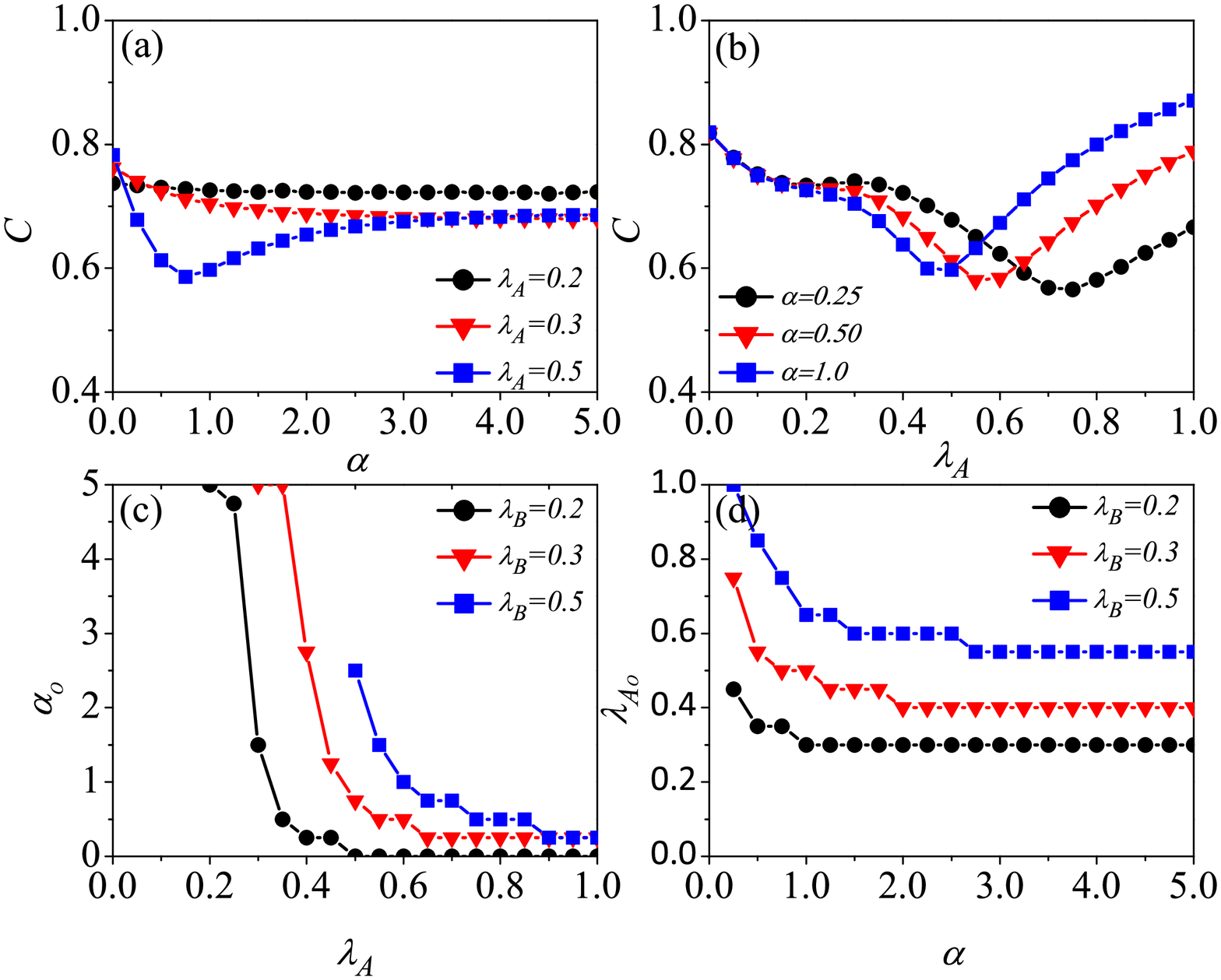,width=1\linewidth}
\renewcommand\thefigure{S\arabic{figure}}
\caption{\textbf{Impacts of social reinforcement effect and information transmission rate on the social cost and the optimal control.} For RR-ER double-layer network, the social cost $C$ versus the parameters of $\alpha$ and $\lambda_A$ in subfigures (a) and (b), respectively. Here the value of $\lambda_B$ is fixed at $0.3$. The optimal $\alpha_o$ versus $\beta_A$ and optimal $\lambda_{Ao}$ versus $\alpha$ in subfigures (c) and (d), respectively. In (a), we select three different values of $\lambda_A$(0.2, 0.3, and 0.5), corresponding to the black circle solid, red triangle solid, and blue square solid lines, respectively. In (b), different values of $\alpha$ (0.25, 0.5 and 1.0) corresponds to the black circle solid, red triangle solid, and blue square solid lines, respectively. (c) the $\alpha_o$ versus $\lambda_A$ and (d) the $\lambda_{Ao}$ versus $\alpha$ under different $\lambda_B$ (0.2, 0.3 and 0.5)  corresponds to the black circle solid, red triangle solid, and blue square solid lines, respectively.}
\label{figS4}
\end{center}
\end{figure}

\begin{figure}
\epsfig{file=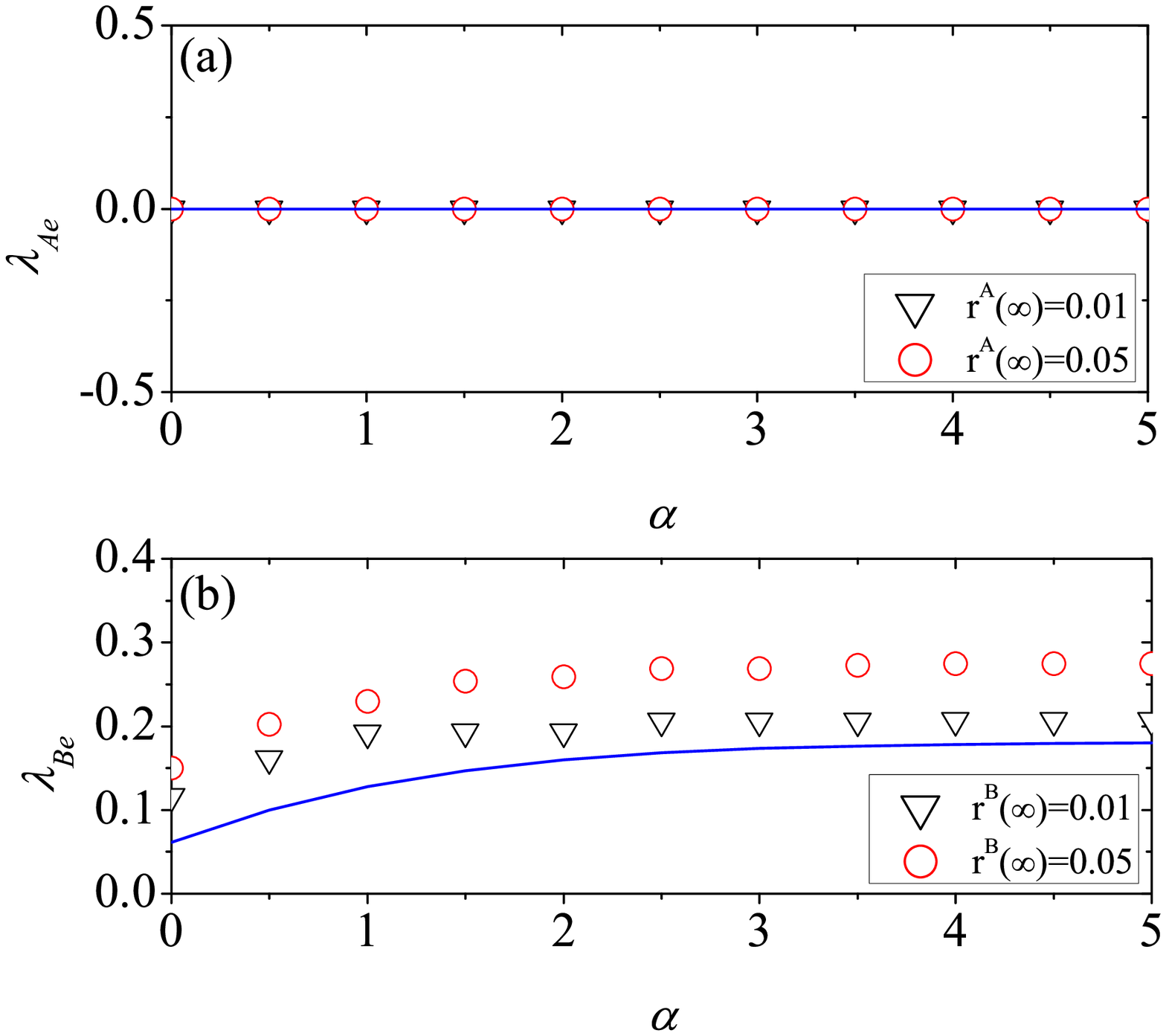,width=0.7\linewidth}
\renewcommand\thefigure{S\arabic{figure}}
\caption{\textbf{The impacts of social reinforcement effect on the outbreak threshold.} For SF-SF double-layer network, the reference information threshold $\lambda_{Ae}$ and the reference epidemic threshold $\lambda_{Be}$ as the function of $\alpha$ are obtained by numerical simulation. Owing to the difficulty of determining the threshold values from numerical predictions, we respectively take the critical density where the final recovery density in layer $A$ ($B$) are 0.01 (black down triangles), and 0.05(red circles) as the reference threshold values. The blue solid line is the corresponding theoretical prediction from Eqs. (S14) and (S16). (a) In communication layer $A$, the reference information threshold $\lambda_{Ae}$ as a function of $\alpha$ when $\lambda_B$ is set as $0.5$; (b) In physical contact layer $B$, the reference epidemic threshold $\lambda_{Be}$ as a function of $\alpha$ at $\lambda_A=0.5$.}
\label{figS5}
\end{figure}

\begin{figure}
\begin{center}
\epsfig{file=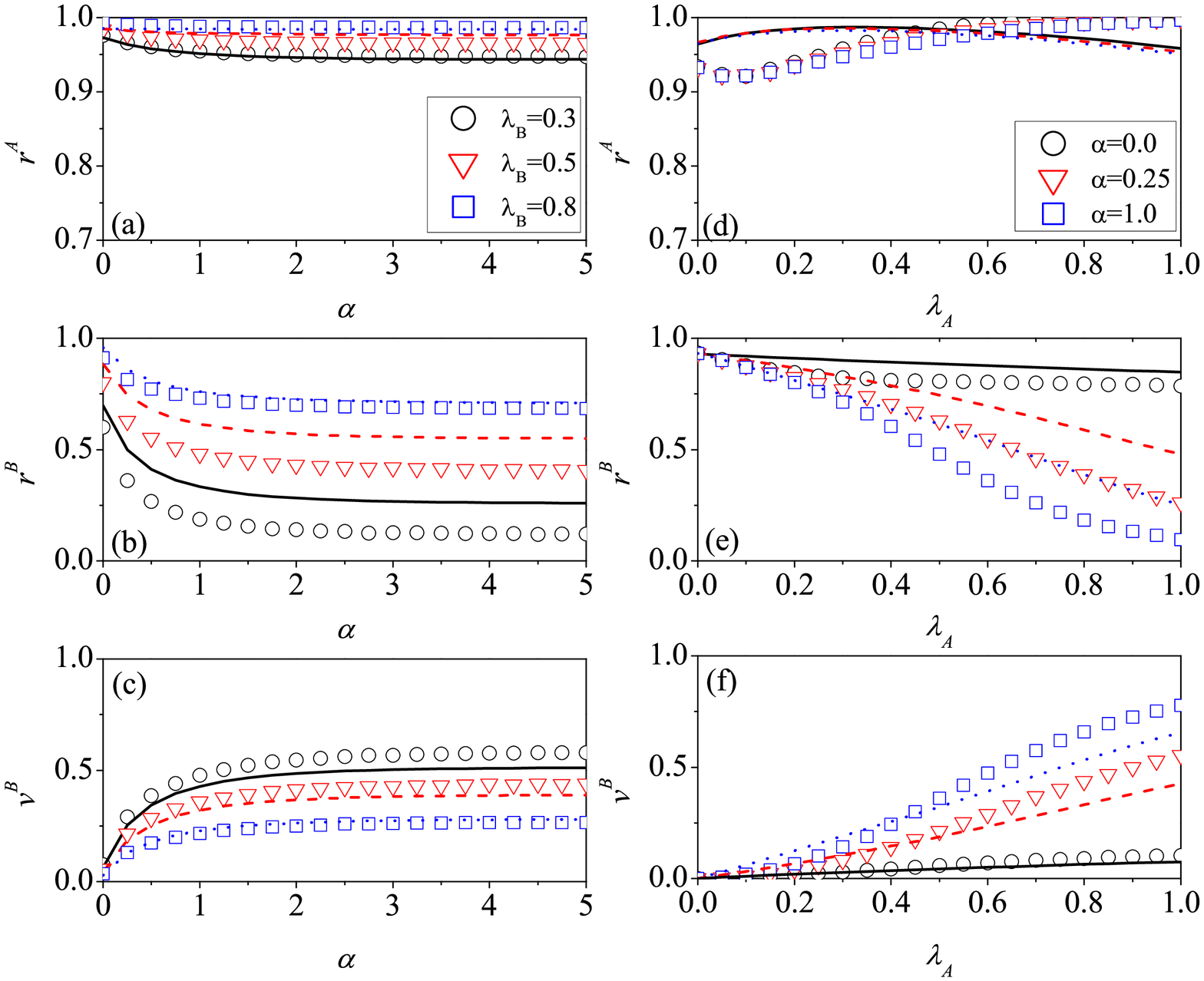,width=1\linewidth}
\renewcommand\thefigure{S\arabic{figure}}
\caption{\textbf{The impacts of social reinforcement effect and information transmission rate on final states.} For SF-SF double-layer network, subfigures (a), (b), and (c) show the values of $r^A$, $r^B$ and $v^B$ as a function of $\alpha$ for different values of $\lambda_B$ (0.3, 0.5, and 0.8), with the analytical predictions corresponding to the black solid, red dashed, and blue doted lines, respectively. When $\lambda_A$ is set as $0.5$. Subfigures (d), (e), and (f) illustrate the values of $r^A$, $r^B$ and $v^B$ versus the parameter $\lambda_A$ for different values of $\alpha$ (0.0, 0.25, and 1.0), corresponding to the black solid, red dashed, and blue doted lines respectively. When $\lambda_B$ is fixed at $0.5$.}
\label{figS2}
\end{center}
\end{figure}

\begin{figure}
\epsfig{file=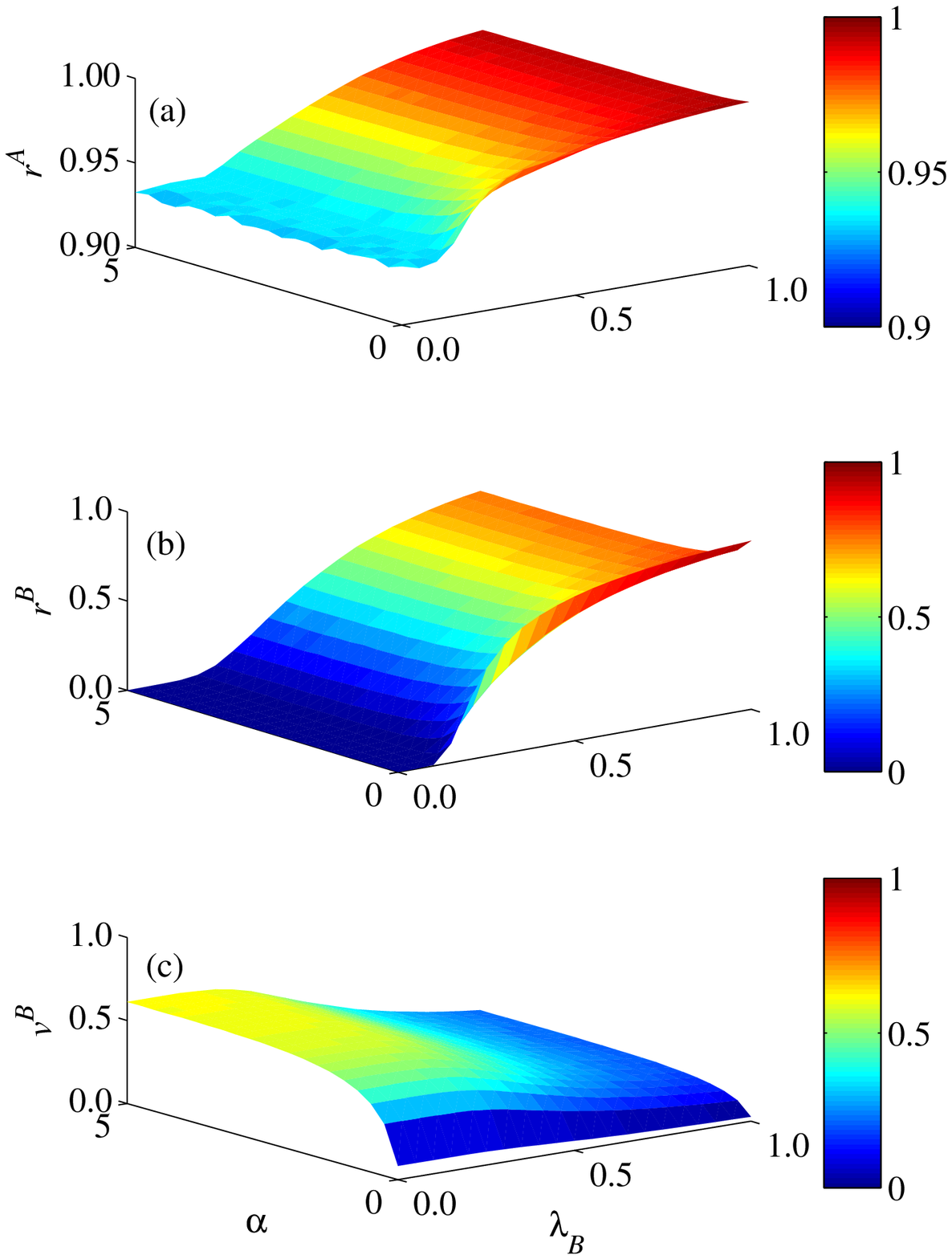,width=0.7\linewidth}
\renewcommand\thefigure{S\arabic{figure}}
\caption{\textbf{A systematic investigation of social reinforcement effect and disease transmission rate impact on final states.} For SF-SF double-layer network, (a) recovered density $r^A$,
(b) recovered density $r^B$, (c) the vaccination density $v^B$ versus $\alpha$ and $\beta_B$ for $\lambda_A=0.5$.}
\label{figS7}
\end{figure}

\begin{figure}
\begin{center}
\epsfig{file=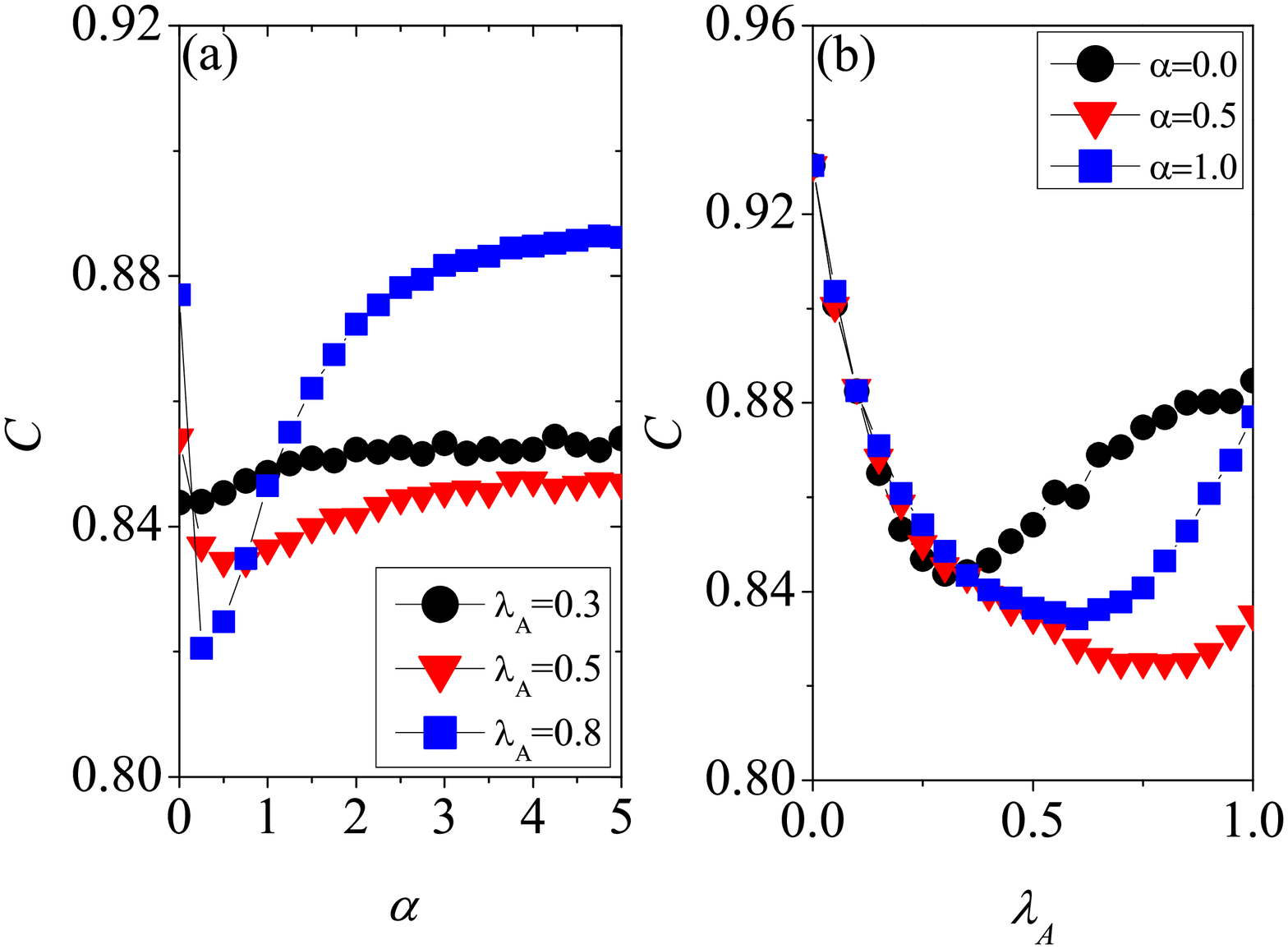,width=0.7\linewidth}
\renewcommand\thefigure{S\arabic{figure}}
\caption{\textbf{Impacts of social reinforcement effect and information transmission rate on the social cost and the optimal control.} For SF-SF double-layer network, the social cost $C$ versus the parameters of $\alpha$ and $\lambda_A$ in subfigures (a) and (b), respectively. Here the value of $\lambda_B$ is fixed at $0.5$. In (a), we select three different values of $\lambda_A$(0.3, 0.5, and 0.8), corresponding to the black circle solid, red triangle solid, and blue square solid lines, respectively. In (b), different values of $\alpha$ (0.0, 0.5 and 1.0) corresponds to the black circle solid, red triangle solid, and blue square solid lines, respectively.}
\label{figS8}
\end{center}
\end{figure}

\begin{figure}
\begin{center}
\epsfig{file=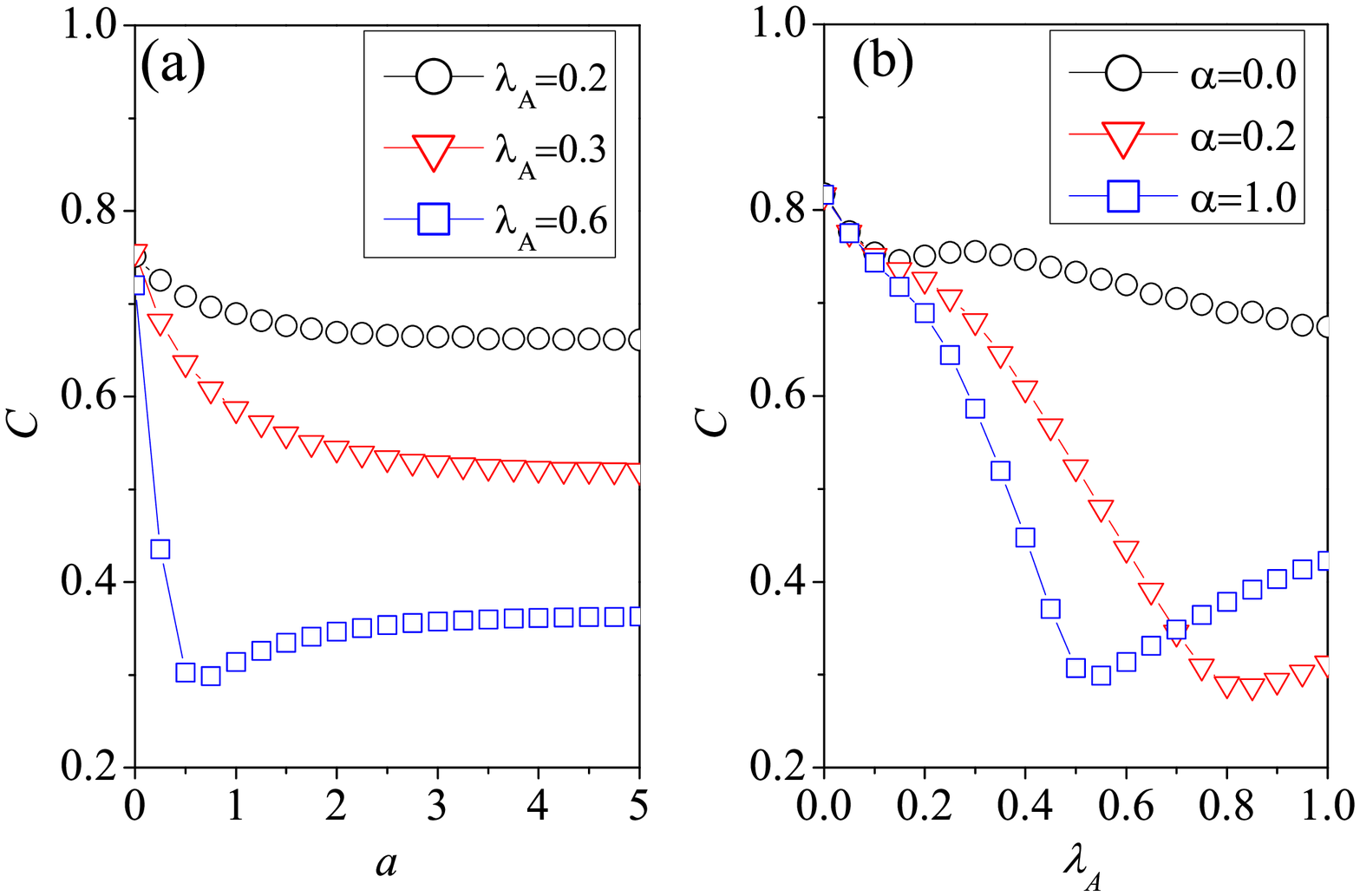,width=0.7\linewidth}
\renewcommand\thefigure{S\arabic{figure}}
\caption{\textbf{Impacts of social reinforcement effect and information transmission rate on the social cost.} For SF-ER double-layer network, the social cost $C$ versus the parameters of $\alpha$ and $\lambda_A$ in subfigures (a) and (b), respectively. Here the value of $\lambda_B$ is fixed at $0.3$. $c_R/c_V=2.$}
\label{figS9}
\end{center}
\end{figure}

\begin{figure}
\begin{center}
\epsfig{file=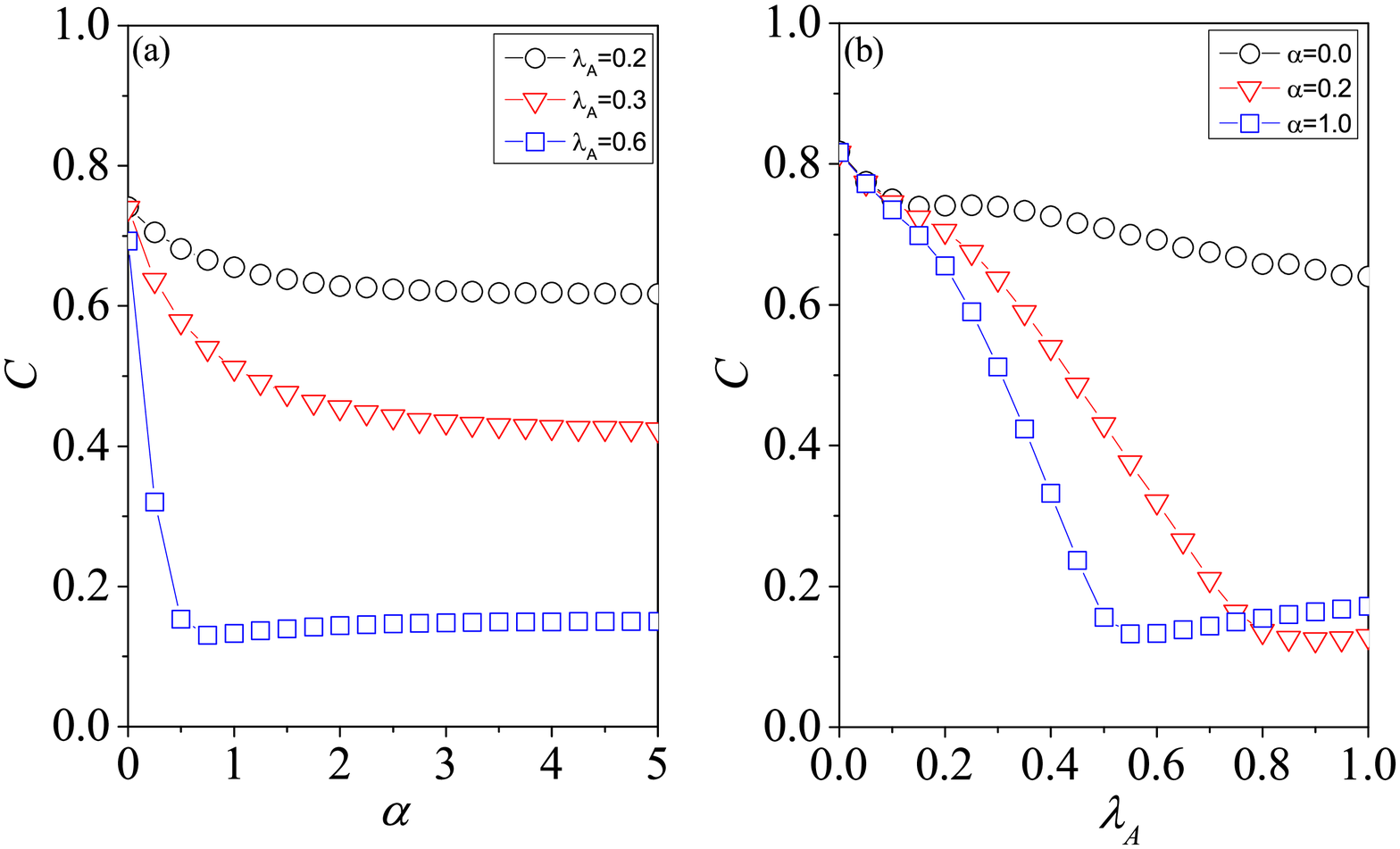,width=0.7\linewidth}
\renewcommand\thefigure{S\arabic{figure}}
\caption{\textbf{Impacts of social reinforcement effect and information transmission rate on the social cost.} For SF-ER double-layer network, the social cost $C$ versus the parameters of $\alpha$ and $\lambda_A$ in subfigures (a) and (b), respectively. Here the value of $\lambda_B$ is fixed at $0.3$. $c_R/c_V=5.$ }
\label{figS10}
\end{center}
\end{figure}


\begin{references}

\bibitem{Anderson:1992}
Anderson, R. M. \& May, R. M. \emph{Infectious diseases of humans}
(Oxford University Press, Oxford, 1991).
\bibitem{Hethcote:2000}
Hethcote, H. W. The mathematics of infectious diseases.
\emph{SIAM review} \textbf{42}, 599-653 (2000).
\bibitem{Daley:2001}
Daley, D. J., Gani, J., \& Gani, J. M. \emph{Epidemic modelling: an introduction}
(Cambrige university press, England, 2001).
\bibitem{Pastor-Satorras:2001}
Pastor-Satorras, R., \& Vespignani, A. Epidemic Spreading in Scale-Free Networks.
\emph{Phys. Rev. Lett.} \textbf{86}, 3200 (2001).
\bibitem{Newman:2002}
Newman, M. E. J. Spread of epidemic disease on networks.
\emph{Phys. Rev. E} \textbf{66}, 016128 (2002).
\bibitem{Moreno:2002}
Moreno, Y., Pastor-Satorras, R. \& Vespignani, A. Epidemic outbreaks in complex
heterogeneous networks.
\emph{Eur. Phys. J. B} \textbf{26}, 521-529 (2002).
\bibitem{Romualdo:2015}
Pastor-Satorras, R., Castellano, C., Mieghem, P. V., Vespignani, A. Epidemic processes in complex networks.
\emph{Rev. Mod. Phys.} \textbf{87}, 925 (2015).
\bibitem{Zanette:2002}
Zanette, D. H. Dynamics of rumor propagation on small-world networks.
\emph{Phys. Rev. E} \textbf{65}, 041908 (2002).
\bibitem{Liu:2003}
Liu, Z., Lai, Y. C. \& Ye, N. Propagation and immunization of infection on general
networks with both homogeneous and heterogeneous components.
\emph{Phys. Rev. E} \textbf{67}, 031911 (2003).
\bibitem{Noh:2004}
Noh, J. D. \& Rieger, H. Random Walks on Complex Networks.
\emph{Phys. Rev. Lett.} \textbf{92}, 118701 (2004).
\bibitem{Kiss:2010}
Kiss, I. Z., Cassell, J., Recker, M. \& Simon, P. L. The impact of information
transmission on epidemic outbreaks.
\emph{Math. Biosci.} \textbf{225}, 1-10 (2010).
\bibitem{Sahneh:2012}
Sahneh, F. D., Chowdhury, F. N. \& Scoglio, C. M. On the existence of a threshold for
preventive behavioral responses to suppress epidemic spreading.
\emph{Sci. Rep.} \textbf{2}, 632 (2012).
\bibitem{Wu:2012}
Wu, Q., Fu, X., Small, M. \& Xu, X. J. The impact of awareness on epidemic
spreading in networks.
\emph{Chaos} \textbf{22}, 013101 (2012).
\bibitem{Ruan:2012}
Ruan, Z., Tang, M. \& Liu, Z. Epidemic spreading with information-driven vaccination.
\emph{Phys. Rev. E} \textbf{86}, 036117 (2012).
\bibitem{Jo:2006}
Jo, H.-H., Baek, S. K. \& Moon, H.-T. Immunization dynamics on a two-layer
network model.
\emph{Physica A} \textbf{361}, 534-542 (2006).
\bibitem{Funk:2010JTB}
Funk, S., Gilad, E. \& Jansen, V. A. A. Endemic disease, awareness, and local
behavioural response.
\emph{J. Theor. Biol} \textbf{264}, 501-509 (2010).
\bibitem{Funk:2009}
Funk, S., Gilad, E., Watkins, C. \& Jansen, V. A. A. The spread of awareness and its impact
on epidemic outbreaks.
\emph{Proc. Natl. Acad. Sci. USA} \textbf{106}, 6872-6877 (2009).
\bibitem{Granell:2013}
Granell, C., G\'{o}mez, S. \& Arenas, A. Dynamical interplay between awareness and epidemic
spreading in multiplex networks.
\emph{Phys. Rev. Lett.} \textbf{111}, 128701 (2013).
\bibitem{Wei:2014}
Wang, W. \emph{et al}. Asymmetrically interacting spreading dynamics on complex layered networks.
\emph{Sci. Rep.} \textbf{4}, 5097 (2014).
\bibitem{Kivel:2014}
Kivel\"{a}, M. \emph{et al}. Multilayer networks.
\emph{Journal of Complex Networks} \textbf{2}, 203-271. (2014).
\bibitem{Kim:2013}
Kim, J. Y. \& Goh, K.-I. Coevolution and Correlated Multiplexity in Multiplex
Networks.
\emph{Phys. Rev. Lett.} \textbf{111}, 058702 (2013).
\bibitem{Boccaletti:2014}
Boccaletti, S. \emph{et al}. The structure and dynamics of multilayer networks.
\emph{Phys. Rep.} \textbf{544}, 1-122 (2014).
\bibitem{Salehi:2014}
Salehi, M. \emph{et al}. Diffusion processes in multilayer networks.
\emph{arXiv e-print} \textbf{1405}, 4329 (2014).
\bibitem{Parshani:2010}
Parshani, R., Rozenblat, C., Ietri, D., Ducruet, C. \& Havlin, S. Inter-similarity
between coupled networks.
\emph{Europhys. Lett.} \textbf{92}, 68002 (2010).
\bibitem{Shao:2011}
Shao, J., Buldyrev, S. V., Havlin, S. \& Stanley, H. E. Cascade of failures in coupled
network systems with multiple support-dependence relations.
\emph{Phys. Rev. E} \textbf{83}, 036116 (2011).
\bibitem{Lee:2012}
Lee, K.-M., Kim, J. Y., Cho, W.-K., Goh, K.-I. \& Kim, I.-M. Correlated multiplexity
and connectivity of multiplex random networks.
\emph{New J. Phys.} \textbf{14}, 033027 (2012).
\bibitem{Zhang:2014}
Zhang, H. F., Xie, J. R., Tang, M., Lai, Y. C. Suppression of epidemic spreading in complex networks by local information based behavioral responses.
\emph{Chaos} \textbf{24}, 043106 (2014).
\bibitem{Young:2011}
Young, H. P. The dynamics of social innovation.
\emph{Proc. Natl. Acad. Sci. USA} \textbf{108}, 21285-21291 (2011).
\bibitem{Centola:2011}
Centola, D. An experimental study of homophily in the adoption of health behavior.
\emph{Science} \textbf{334}, 1269-1272 (2011).
\bibitem{Centola:2010}
Centola, D. The spread of behavior in an online social network experiment.
\emph{Science} \textbf{329}, 1194-1197 (2010).
\bibitem{Dodds:2004}
Dodds, P. S., Watts, D. J. Universal behavior in a generalized model of contagion.
\emph{Phys. Rev. Lett.} \textbf{92}, 218701 (2004).
\bibitem{Dodds:2005}
Dodds, P. S., Watts, D. J. A generalized model of social and biological contagion.
\emph{J. Theor. Biol.} \textbf{232}, 587-604 (2005).
\bibitem{Weiss:2014}
Weiss, C. H. \emph{et al}. Adoption of a high-impact innovation in a homogeneous population.
\emph{Phys. Rev. X} \textbf{4}, 041008 (2014).
\bibitem{Centola:2007}
Centola, D., Macy, M. Complex contagions and the weakness of long ties.
\emph{American Journal of Sociology} \textbf{113}, 702-734 (2007).
\bibitem{Zhang:2013}
Zhang, J., Liu, B., Tang, J., Chen, T. \& Li, J. Social influence locality for modeling retweeting behaviors.
\emph{Proceedings of the Twenty-Three International Joint Conference on Artificial Intelligence} Beijing, China. Menlo Park, California, USA: AAAI Press. (2013).
\bibitem{Hodas:2014}
Hodas, N. O. \& Lerman, K. The simple rules of social contagion.
\emph{Sci. Rep.} \textbf{4}, 4343 (2014).
\bibitem{ZhangHF:2014}
Zhang, H. F., Wu, Z. X., Tang, M. \& Lai, Y. C. Effects of behavioral response and vaccination policy on epidemic spreading-an approach
based on evolutionary-game dynamics. \emph{Sci. Rep.} \textbf{4}, 5666 (2014).
\bibitem{Altarelli:2014}
Altarelli, F., Braunstein, A., Dall¡¯Asta, L., Wakeling, J. R. \& Zecchina, R. Containing epidemic outbreaks by
message-passing techniques. \emph{Phys. Rev. X} \textbf{4}, 021024 (2014).
\bibitem{Marceau:2011}
Marceau, V., No\"{e}l, P.-A., H\'{e}bert-Dufresne, L., Allard, A. \& Dub\'{e}, L. J.
Modeling the dynamical interaction between epidemics on overlay networks.
\emph{Phys. Rev. E} \textbf{84}, 026105 (2011).
\bibitem{Wang:2015}
Wang, W., Tang, M., Zhang, H. F., Lai, Y. C. Dynamics of social contagions with memory of nonredundant information.
\emph{Phys. Rev. E} \textbf{92}, 012820 (2015).
\bibitem{Karrer:2011}
Karrer, B. \& Newman, M. E. J. Competing epidemics on complex networks.
\emph{Phys. Rev. E} \textbf{84}, 036106 (2011).
\bibitem{Newman:2005}
Newman, M. E. J. Power laws, Pareto distributions and Zipf's law.
\emph{Contemp. Phys.} \textbf{46}, 323¨C351 (2005).
\bibitem{Catanzaro:2005}
Catanzaro, M., Bogu\~{n}\'{a} M. \& Pastor-Satorras, R. Generation of uncorrelated random scale-free networks.
\emph{Phys. Rev. E} \textbf{71}, 027103 (2005).

\bibitem{Erdos:1959}
Erd\H{o}s, P. \&  R\'{e}nyi. On random graphs.
\emph{Publ. Math.} \textbf{6}, 290-297 (1959).

\bibitem{Shu:2015}
Shu, P., Wang, W., Tang, M., Do, Y. Numerical identification of epidemic thresholds for susceptible-infected-recovered model on finite-size networks. \emph{Chaos} \textbf{25}, 063104 (2015).
\bibitem{Argollo:2004}
Argollo, M., Barab\'{a}si, A.-L. Separating Internal and External Dynamics of Complex Systems. \emph{Phys. Rev. Lett.} \textbf{93}, 068701 (2004).


\bibitem{Ferreri:2014}
Ferreri, L., Bajardi, P., Giacobini, M., Perazzo, S. \& Venturino, E. Interplay of network dynamics and heterogeneity of ties on spreading
dynamics. \emph{Phys. Rev. E} \textbf{90}, 012812 (2014).
\bibitem{Stohr:2004}
St{\"o}hr, K., \& Esveld, M. Will vaccines be available for the next influenza pandemic? \emph{Science} \textbf{306}, 2195-2196 (2004).
\bibitem{Reluga:2006}
Reluga, T. C., Bauch, C. T. \& Galvani, A. P. Evolving public perceptions and stability in vaccine uptake. \emph{Math. Biosci.} \textbf{204}, 185-198 (2006).
\bibitem{Mbah:2012}
Ndeffo Mbah, M. L. \emph{et al}. The impact of imitation on vaccination behavior in social contact networks.
\emph{PLoS Comput. Biol.} \textbf{8}, e1002469 (2012).
\bibitem{Dybies:2004}
Dybiec B., Kleczkowski A., Gilligan C. Controlling disease spread on networks with incomplete knowledge. \emph{Phys. Rev. E} \textbf{70}, 066145(2004).
\bibitem{Kleczkowski:2006}
Kleczkowski A., Dybiec B., Gilligan C. A. Economic and social factors in designing disease control strategies for epidemics on networks.
\emph{Acta Phys. Pol. B} \textbf{37}, 3017-3026(2006).
\bibitem{Meltzer:1999}
Meltzer M., Cox N., Fukuda K. The economic impact of pandemic influenza in the United States: priorities for intervention. \emph{Emerg. Infect. Dis.} \textbf{5}, 659(1999).
\bibitem{Weycker:2004}
Weycker D. \emph{et al}. Population-wide benefits of routine vaccination of children against influenza. \emph{Vaccine} 23, 1284-1293(2004).
\bibitem{Galvani:2007}
Galvani, A. P., Reluga, T. C., and Chapman, G. Long-standing influenza vaccination policy is in accord with individual self-interest but not with the utilitarian optimum. \emph{Proc. Natl. Acad. Sci. USA} \textbf{104}, 5692 (2007).
\bibitem{Dybies:2004}
Dybiec, B. Controlling disease spread on networks with incomplete knowledge. \emph{Phys. Rev. E} \textbf{70}, 066145 (2004).
\bibitem{Kleczkowski:2012}
Kleczkowski, A., Ole, K., Gudowska-Nowak, E. \& Gilligan, C. A. Searching for
the most cost-effective strategy for controlling epidemics spreading on regular
and small-world networks. \emph{J. R. Soc. Interface} \textbf{9}, 158-169(2012).
\bibitem{Bauch:2013}
Bauch, C. T. \& Galvani, A. P. Social Factors in Epidemiology.
\emph{Science} \textbf{342}, 47-49 (2013).
\bibitem{Newman:2010}
Newman, M. E. J. \emph{Networks An Introduction}
(Oxford University Press, Oxford, 2010).
\end{references}

\begin{references}
\bibitem{sBarthelemy11:2004}
Barth\'{e}lemy, M.,   Barrat, A.,  Pastor-Satorras, R. \&  Vespignani, A. Velocity
 and Hierarchical Spread of Epidemic Outbreaks in Scale-Free Networks. \emph{Phys. Rev. Lett.} \textbf{92}, 178701 (2004).
\bibitem{sNewman:2010}
Newman, M. E. J. \emph{Networks An Introduction}
(Oxford University Press, Oxford, 2010).
\bibitem{sSaumell:2012}
Saumell-Mendiola, A., Serrano, M. {\'A}. \& Bogu{\~n}{\'a}, M. Epidemic spreading on interconnected networks.
\emph{Phys. Rev. E} \textbf{86}, 026106 (2012).
\bibitem{sMieghem:2011}
Mieghem, P. V.
\emph{Graph Spectra for Complex Networks} (Cambrige university press, England, 2011).
\bibitem{Karrer:2011}
Karrer, B. \& Newman, M. E. J. Competing epidemics on complex networks.
\emph{Phys. Rev. E} \textbf{84}, 036106 (2011).
\bibitem{sWei:2014}
Wang, W. \emph{et al}. Asymmetrically interacting spreading dynamics on complex layered networks.
\emph{Sci. Rep.} \textbf{4}, 5097 (2014).
\bibitem{sMarro:1999}
Marro, J. \& Dickman, R. \emph{Nonequilibrium Phase Transitions in Lattice Models}
(Cambrige university press, Cambrige, 1999).
\bibitem{sErdos:1959}
Erd\H{o}s, P. \&  R\'{e}nyi. On random graphs. \emph{Publ. Math.} \textbf{6}, 290-297 (1959).
\bibitem{sNewman:2001-2}
Newman, M. E. J., Strogatz, S. H. \& Watts, D. J.
Random graphs with arbitrary degree distributions and
their applications. \emph{Phys. Rev. E} \textbf{64}, 026118 (2001).
\bibitem{sNewman:2005-2}
Newman, M. E. J. Power laws,
Pareto distributions and Zipf's law.
\emph{Contemp. Phys.} \textbf{46}, 323-351 (2005).
\bibitem{sCatanzaro:2005}
Catanzaro, M., Bogu\~{n}\'{a}, M. \& Pastor-Satorras, R.
Generation of uncorrelated random scale-free networks.
\emph{Phys. Rev. E} \textbf{71}, 027103 (2005).



\end{references}
\end{document}